\documentclass[10pt,journal,compsoc]{IEEEtran}

\usepackage{xurl}
\usepackage{cite}  
\usepackage{amsmath}
\usepackage{amstext}
\usepackage{amssymb}
\usepackage{amsfonts}
\usepackage{tikz}
\usepackage{amsthm}
\usepackage{algorithmic}
\usepackage{graphicx}
\usepackage{textcomp}
\usepackage{xcolor}
\usepackage{hhline}
\usepackage{listings}
\usepackage{url}
\usepackage{hyperref}
\usepackage{xspace}
\usepackage{multicol} %package to balance the references page

\definecolor{commentgreen}{RGB}{0,153,76}
\definecolor{light-gray}{gray}{0.85}
\theoremstyle{plain}

\newcommand{\name}{\textsc{CoCoA}\xspace}

\newcommand{\LONG}[1]
{{}}
\begin{document}

\title{\Large Detecting Vulnerabilities in Encrypted Software Code while Ensuring Code Privacy}

\author{
    \IEEEauthorblockN{Jorge Martins}
    \IEEEauthorblockA{LASIGE, DI, Faculdade de Ci\^encias\\ Universidade de Lisboa, Portugal\\ fc51033@alunos.fc.ul.pt}
    \and
    \IEEEauthorblockN{David Dantas}
    \IEEEauthorblockA{LASIGE, DI, Faculdade de Ci\^encias\\ Universidade de Lisboa, Portugal\\ fc56331@alunos.fc.ul.pt}
    \and
    \IEEEauthorblockN{Rafael Ramires}
    \IEEEauthorblockA{LASIGE, DI, Faculdade de Ci\^encias\\ Universidade de Lisboa, Portugal\\ rframires@ciencias.ulisboa.pt}
    \and
    \IEEEauthorblockN{Bernardo Ferreira}
    \IEEEauthorblockA{LASIGE, DI, Faculdade de Ci\^encias\\ Universidade de Lisboa, Portugal\\ blferreira@ciencias.ulisboa.pt}
    \and
	\IEEEauthorblockN{Ib\'{e}ria Medeiros}
	\IEEEauthorblockA{LASIGE, DI, Faculdade de Ci\^{e}ncias,\\Universidade de Lisboa - Portugal\\ ivmedeiros@fc.ul.pt}
}

\author{Jorge Martins, David Dantas, Rafael Ramires, 
    Bernardo Ferreira,~\IEEEmembership{Member,~IEEE,}
    Ib\'{e}ria~Medeiros,~\IEEEmembership{Member,~IEEE}% <-this % stops a space
	\IEEEcompsocitemizethanks{\IEEEcompsocthanksitem J. Martins, D. Dantas, R. Ramires, B. Ferreira and I. Medeiros are with the LASIGE, DI, Faculdade de Ci\^{e}ncias, Universidade de Lisboa - Portugal (e-mail: fc51033@alunos.fc.ul.pt, fc56331@alunos.fc.ul.pt, rframires@ciencias.ulisboa.pt, blferreira@ciencias.ulisboa.pt, and ivmedeiros@ciencias.ulisboa.pt).}% <-this % 
}

\markboth{IEEE TRANSACTIONS ON DEPENDABLE AND SECURE COMPUTING}%
{Shell \MakeLowercase{\emph{et al.}}: Bare Demo of IEEEtran.cls for Computer Society Journals}

%remove the 2 lines to remove page numbers:
\thispagestyle{plain}
\pagestyle{plain}

\IEEEtitleabstractindextext{%
\begin{abstract}
Software vulnerabilities continue to be the primary cause of cyberattacks, largely due to the widespread adoption of web applications, developed for access and management services, that are made available without proper testing. It is crucial to identify vulnerabilities in applications' source code before attackers gain access to them and exploit any vulnerability they may contain.
Developers have used static analysis tools to find vulnerabilities in unprotected application code, and software testing companies have started offering software code analysis as a service to assist developers in these findings. Such services require access to unprotected code, which raises concerns about its privacy and intellectual property theft. Moreover, attackers can also perform this analysis using similar tools, if they gain access to the code.
It is, therefore, beneficial to have a system that can maintain code privacy by protecting it with cryptographic techniques, while still allowing authorised people to detect vulnerabilities in the encrypted code. 
This paper presents such a solution, a novel approach to \emph{Software Quality and Privacy} that allows source code to be analysed in a protected manner, preserving its privacy.
The proposed solution combines Static Analysis with Searchable Symmetric Encryption (SSE) for confidential vulnerability detection, enabling data and dependency tracking for data flow analysis over encrypted source code. The solution represents the code's data and control flows as an \emph{Encrypted Inverted Index}, in a connected way that enables SSE's queries for vulnerability discovery.
The solution was implemented as the \name tool and evaluated experimentally with synthetic and real PHP web applications. Results show that \name has similar precision as standard (non-confidential) static analysis tools -- 93\% -- with real applications, requiring only 209 ms to process 4k lines of code - a modest overhead of 42.7\% compared to a non-confidential baseline.
With this novel approach, this paper also defines a new research field -- \emph{Confidential Code Analysis} --, from which other types of code analysis tasks and approaches can be derived. We believe this work paves the way for this promising area of research, from which \emph{Software Quality and Privacy} can greatly benefit.
\end{abstract}

% Note that keywords are not normally used for peer-reviewed papers.
 }

%
% This command processes the author and affiliation and title
% information and builds the first part of the formatted document.
\maketitle
\IEEEdisplaynontitleabstractindextext

%\input{sections/1-introduction}
%%%%%%%%%%%%%%%%%%%%%%%%%%%%%%%
\section{Introduction}

\IEEEPARstart{S}oftware has played an essential role in our global technological world.
Despite the efforts software companies have made to build quality software, the number of vulnerabilities continues to increase~\cite{cve-date:23}, so much so that the number of cyber attacks has risen to the highest record in recent years~\cite{CheckPoint:24}. 
This increase is justified by the accelerating digital transformation, and indicates that inadequately tested software has been made available to the public, thus causing a decline in overall quality~\cite{Daley:17}.
Web applications are at the root of this transformation, serving as the primary means of accessing and managing services; hence, the vulnerabilities they contain are a source for this security problem.
This makes these attractive targets for attackers~\cite{Veracode:23}, with SQL injection (SQLi) and cross-site scripting (XSS) being among the most exploited vulnerabilities~\cite{owasp}. 

To contradict these figures, static analysis tools~\cite{pixy,wap,Morgado:20} have been used to analyse unprotected source code, identifying vulnerabilities and suggesting how to fix them. 
Developers and software testing companies have resorted to these tools to achieve software quality assurance~\cite{SQA:18,Lee:14,anacov}, with the latter also starting to offer them as a cloud-based service~\cite{scnsoft,qamentor,veracode:24}. 
In this service, software can undergo a series of code analysis tasks, including vulnerability detection, to properly test the software and improve its quality and reliability~\cite{staticviolations,huang2,last,stat,Chess:04}.
However, this requires access to the unprotected source code, raising concerns about its privacy and theft of intellectual property~\cite{wired22source}. 
Moreover, attackers can also perform this analysis using similar tools if they gain access to the code.

Code obfuscation~\cite{Schrittwieser:16,Wu:16,Banescu:18,Hosseinzadeh:18,Kang:21} and code encryption~\cite{Cho:11,Cappaert:06,Dong:16} techniques have the goal of making the code unreadable. They are thereby seen as a way for software companies to protect their code against potential copyright violations, reverse engineering, and code privacy preservation, thus preventing (malicious) third parties from inspecting it.
When applied, they can transform (i.e., obfuscate/encrypt) the entire code or specific portions (e.g., entry points, sensitive sinks), making it totally or partially illegible for reading. As such, the level of protection offered by these techniques relies on the strength of the algorithms and cryptographic keys used. We have observed that such algorithms have been broken~\cite{Hosseinzadeh:18,Cappaert:06} and cryptographic keys have been discovered~\cite{Cho:11}, compromising code protection. 
Moreover, when the code needs to be analysed (even by developers), it is unprotected (de-obfuscated or decrypted) during the analysis process, which can compromise its secrecy. Hence, although such techniques provide a sense of protection, they do not offer real guarantees of code privacy, and are impractical for third parties to conduct code analysis tasks.
An interesting solution to overcome these concerns would be, therefore, to have a system that guarantees the confidentiality of the code by protecting it with cryptographic techniques, while at the same time allowing authorised people to perform code analysis tasks on the encrypted code, such as detecting vulnerabilities.

This paper presents such a solution, a novel approach that protects software code by using an encryption scheme that enables vulnerability finding on the encrypted code without revealing its contents during the analysis.
The approach combines \textit{Static Code Analysis} (SCA)~\cite{Morgado:20} with \textit{Searchable Symmetric Encryption} (SSE)~\cite{praticalsearchable}, two disjoint research fields that would seem unlikely to be combined. 
In SCA, the application's code is represented by code property graphs (e.g., control flow graph and data flow graph)~\cite{ferrante,yamaguchi,eff} which can then be used for different purposes, such as vulnerability detection~\cite{rips}. Static analysis tools navigate them, looking for what they were programmed for. 
They may employ various SCA techniques, with taint analysis~\cite{pixy} being the most commonly applied, which allows tracking entry points (user inputs like \texttt{\$\_POST} in PHP) and checking if they reach sensitive sinks (e.g., \texttt{echo} for XSS). 
However, these tools require access to the application's unprotected code, without it being obfuscated nor encrypted.
In turn, in SSE~\cite{curtmola2006searchable}, text documents are represented through an inverted index~\cite{manning2008introduction}, where keywords (i.e., the documents' most relevant words) are mapped to the documents where they appear. This index is encrypted in a manner that it leaks no information when stored in remote (untrusted) servers, but nonetheless, authorised clients can query it to find index positions (i.e., the documents) containing the keywords they search for.

We propose combining these two research fields to enable \emph{static analysis capable of operating on and understanding encrypted code}, and implement the approach as a new tool named \name (\emph{\textsc{Co}nfidential dete\textsc{C}ti\textsc{o}n of vulner\textsc{A}bilities}).
\name\ processes the source code of the developer's application, extracting its data and control flows and building the \emph{Data and Control Flow Graph} (DCFG -- a type of program dependence graph~\cite{ferrante}). Then it represents the DCFG as an inverted index and encrypts it similarly to SSE, resulting in an \textit{Encrypted Inverted Index} (EII).
Subsequently, the EII (i.e., the encrypted DCFG) is used by authorised people to perform static code analysis tasks and find vulnerabilities with confidentiality, by querying the index for specific vulnerability classes. In turn, the queries trigger data flow analysis considering the control flows present in the DCFG.

\name can be used in three distinct scenarios, all with the same purpose -- \emph{Software Quality and Privacy} -- i.e., when being used, it ensures code privacy and the developer's intellectual property by encrypting the code, while simultaneously allowing its encrypted analysis, which improves the software's quality by identifying vulnerabilities. 
In the first scenario, developers can use \name only to protect their code and prevent unintended access to it, i.e., without static analysis features. On the second, they can use \name both to protect and test the code against vulnerabilities. Lastly, software testing companies can use \name to offer \emph{Confidential Code Analysis as a Service}. Such a service will, on one hand, allow developers to outsource security analysis tasks to specialised third parties without sacrificing their intellectual property nor the privacy of their code. On the other hand, it will allow testing companies to execute such analysis tasks \emph{out-of-house} rather than the more common yet resource-intensive \emph{in-house} manner (i.e., locally at the developers' companies).
This last scenario is particularly interesting, as it will allow software analysers from testing companies to perform code analysis tasks on encrypted software code, while preserving software privacy.
The idea is that analysers should only access an encrypted version of the developer's code and should learn only the result of their code analysis task (e.g., if the code contains vulnerabilities) and nothing else about the code. In contrast, developers should be able to receive and understand the results of the analysis task and take action accordingly (e.g., remove the identified vulnerabilities).

%% CCA
Furthermore, with \name and the proposed approach, this paper also initiates and defines a new research field -- \emph{Confidential Code Analysis} (CCA) --, which generalises the above-mentioned idea.
CCA can open way for many new and interesting applications, such as different code analysis tasks and other approaches for vulnerability detection: type checking~\cite{Zampetti:17}, code smell verification~\cite{Pecorelli:22}, and fault detection~\cite{Medeiros:22,Pearson:17,Chatterjee:21} are some examples, as well as integration into  software development lifecycle and DevSecOps~\cite{Rajapakse:21}.

%Evaluation
We formally analyse \name's security and experimentally evaluate its effectiveness and performance when used to detect vulnerabilities in PHP web applications, the language used in 77\% of such applications~\cite{WTechs:23}.
In our experiments, we use \name to find SQLi and XSS vulnerabilities in two datasets: one with 3,205 programs from NIST SARD~\cite{samate} and the other with 7 open-source applications.
In both datasets, results show similar precision compared with non-confidential tools -- 80\% and 93\% respectively (see Section~\ref{sec:impl_eval} for details) --,
and an average increase of 42.7\% in performance costs, requiring only 209 ms to process 4k lines of code. We find this to be a modest overhead considering the benefits gained in privacy and functionality. 

%Contributions
In summary, our main contributions are:
(1) the definition of \emph{Confidential Code Analysis} (CCA), a new field of research for \emph{Software Quality and Privacy}, where static code analysis tasks are performed while preserving code's privacy and assuring its quality; 
(2) a first CCA solution based on SCA and SSE. The solution includes a data structure definition, encryption schemes, and an SSE-based static analysis algorithm for data flow analysis with control flow to represent the DCFG of a program's source code and ensure that SSE queries correctly follow the data dependencies of the code;
(3) \name, a prototype implementation of the solution, tailored for SQLi and XSS vulnerability detection in PHP web applications.
This implementation is open source and available at~\cite{CoCoA:23};
(4) a security analysis and experimental evaluation of \name\, which demonstrate its security, effectiveness, and performance, particularly when used for detecting web vulnerabilities in PHP code.

The remaining of the paper is organised as follows: Section~\ref{sec:CCA} defines CCA. Section~\ref{sec:cocoa} presents an overview of \name and its challenges, while Sections~\ref{sec:ITL} and~\ref{sec:Det_Vuln} detail \name's design, Section~\ref{sec:security} its security analysis, and Section~\ref{sec:Implment} its implementation and deployment scenarios. Section~\ref{sec:impl_eval} evaluates \name in terms of theoretical cost analysis, effectiveness when applied to vulnerability detection, and performance and storage overheads. Section~\ref{app:extensions} shows how the tool can be extended for other scenarios. Lastly, Section~\ref{sec:RW} presents the related work, and Section~\ref{sec:conc} concludes the paper.

%\input{sections/2-CCA}
%%%%%%%%%%%%%%%%%%%%%%%%%%%%%%%
\section{Confidential Code Analysis}
\label{sec:CCA}

%-----------
\subsection{Formal Definition}
We start by formally defining and analysing the concept of \emph{Confidential Code Analysis} (CCA). Our intuition behind this concept is twofold: $i)$ developers should be able to produce an encrypted representation of their code that does not reveal anything by itself, meaning that it can be securely shared with analysers or any other untrusted third party; $ii)$ an analyser can submit a specification for an analysis task it wants to perform and receive (if authorized by the developer) a cryptographic token that, when combined with the encrypted code, outputs the result of the task.

We believe this intuition captures the goal of CCA.
When a developer encrypts its code, this encryption does not reveal anything to adversaries, as in standard encryption. Nonetheless, at a later time, analysers can use this encryption to perform an analysis task of their choice. This step requires additional authorisation in the form of a cryptographic token that the developer must provide to allow the specified analysis. 
This puts the burden of scrutinising analysis requests on the developers, however it allows them to control who can analyse their code and which tasks they can perform. It also ensures that even if the encrypted code falls into the hands of an adversary, without access to one of these tokens it will not learn anything about the code.
One could also think of other design choices, such as having the developer specify a privacy parameter and the system automatically determine which analysis requests to authorise based on that, but we leave the exploration of such ideas for future work.

The next definition captures the intuition described. Formally, a CCA scheme $\prod = (\mathsf{Encrypt}$, $\mathsf{Authorise}$, $\mathsf{Analyse})$ consists of three protocols that are executed between a developer and an analyser:

\begin{itemize}
	\item 
	$\{k,E_k(c)\} \gets \prod.\mathsf{Encrypt}(\lambda,c)$ is executed by the developer and encrypts a software's source code, with inputs security parameter $\lambda$ and source code $c$.
	At the end, the developer produces secret key $k$, which it keeps to itself, and an encrypted representation of the code $E_k(c)$, which can be shared with the analyser.
	
	\item 
	$\{k_f\} \gets \prod.\mathsf{Authorise}(k,f)$ is executed by the developer and creates an authorisation token for a specific analysis task provided by the analyser, with inputs secret key $k$ and the task $f$. At the end of the protocol, the developer produces an authorisation token $k_f$, sharing it with the analyser.
	
	\item 
	$\{f_{k_f}(E_k(c))\} \gets \prod.\mathsf{Analyse}(f,k_f,E_k(c))$ is executed by the analyser and performs analysis task $f$ by evaluating it on the encrypted code $E_k(c)$ through authorisation token $k_f$, i.e., $f(c) = f_{k_f}(E_k(c))$. At the end, the analyser produces the encrypted results of task $f$, which can be shared with the developer.
\end{itemize}

Regarding security, the intuition is that: $E_k(c)$ should not reveal anything about $c$ (\textit{code secrecy}); $f_{k_f}(E_k(c))$ should only reveal the result of task $f$ when applied to $c$ (\textit{analysis secrecy with public result} -- meaning that the 
analyser learns the result of the task (e.g., there is an XSS vulnerability) but nothing else about the code); and $f_{k_f}(E_k(c))$ should output the same result as $f(c)$ (\textit{correctness}).

\subsection{Adversary Model}
We focus on securing \emph{Software Code Analysis as a Service} application scenarios, where the developer's encrypted code is stored in a remote server (e.g., analysers' infrastructure or the cloud), and analysers access it to carry out their code analysis tasks. In this scenario, a semi-honest adaptive adversary is considered, meaning that it will execute the protocol as expected, but may try to break the confidentiality of the code being analysed or learn sensitive information about it that the developer did not explicitly authorise. The adversary may be the analyser itself or a third party impersonating it, or eavesdropping on their communications. We consider the developer to be trusted, as it is its source code that is being protected. We note that this adversary model is reminiscent of the one typically considered in cloud-based applications~\cite{armbrust2010view}.

%\input{sections/3-Overview}
%%%%%%%%%%%%%%%%%%%%%%%%
\section{\name Challenges and Overview}
\label{sec:cocoa}

We aim to create a system that follows the CCA definition and so is capable of analysing encrypted code to discover vulnerabilities. We call it \name and it combines \textit{Static Code Analysis} (SCA) with \textit{Searchable Symmetric Encryption} (SSE). Before introducing our approach, we present the challenges we faced in building such a CCA scheme.

%-------------
\subsection{Challenges}
\label{sec:challenges}

\textbf{\emph{1) How to protect code privacy while supporting analysis tasks?}}
The easiest way to protect the privacy of a program is to encrypt its source code. However, code analysis needs to understand the program's logic, semantics, and control flow, as well as the data dependencies between variables and function calls, which is impossible to do in encrypted code. 
Our idea is to represent the source code in a lightweight intermediate language that preserves these properties and, from there, extract its \emph{Data and Control Flow Graph} (DCFG) and encrypt it with a scheme that allows computations over encrypted data. 
This will allow data flow analysis considering both data flow and control flow information.

\vspace{1mm}
\noindent
\textbf{\emph{2) Which cryptographic schemes can support DCFG analysis?}}
There are many encryption schemes that support computations on encrypted data. Fully Homomorphic Encryption (FHE)~\cite{gentry2009fully} allows additions and multiplications of encrypted values. Since any computation can be expressed as an arithmetic circuit, FHE can be used as a general-purpose tool for encrypted data processing.
Multi-Party Computation~\cite{cramer2015secure} also supports generic encrypted computations, although requiring multiple non-colluding parties. However, both schemes are still far from practical in terms of performance. 
Instead, we propose using SSE~\cite{ferreira2020boolean}, a more practical solution designed for securely storing and searching encrypted text databases in remote untrusted servers. Our idea is to encrypt the DCFG with SSE, but with the necessary adjustments to support software code analysis. 

\vspace{1mm}
\noindent
\textbf{\emph{3) How can we adapt SSE to encrypted software code analysis?}}
Moving from keyword queries in encrypted databases to code analysis in encrypted code seems promising, but it is far from trivial. To accomplish this, many challenges arise, including: 
$i)$ what type of data structure can represent the DCFG, i.e., the code and its data and control flows, for encryption? 
$ii)$ how to encrypt such a data structure so that it does not leak anything about the code when stored? $iii)$ how to perform code analysis tasks through this encrypted data structure, revealing just the result of the intended computation?
We will answer these questions in the next sections as we explain the details of our solution.

%------------
\subsection{Overview}

\begin{figure}[t]
	\centering
	\includegraphics[width=1\columnwidth]{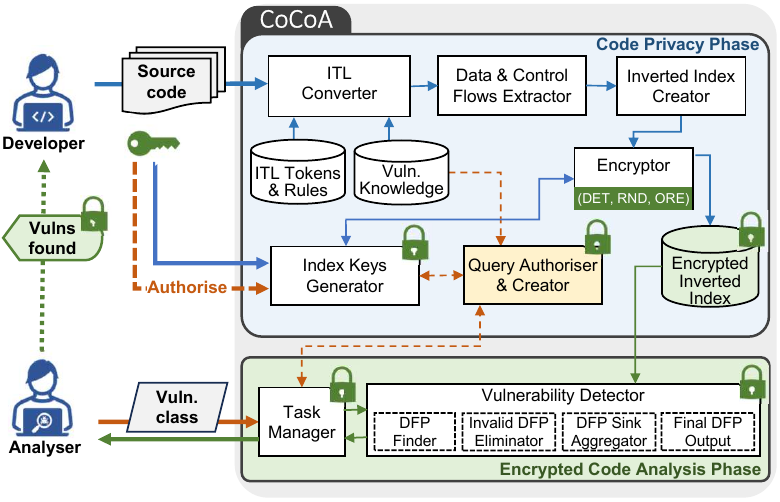}
	%\vspace{-8mm}
	\caption{\name's architecture. ITL stands for \emph{Intermediary Token Language}, DFP for \emph{Data Flow Path}, and Vuln. for \emph{Vulnerability}.}
	\label{fig:arch}
	%\vspace{-4mm} 
\end{figure}

This section presents an overview of \name, the solution that allowed us to solve the previous challenges. As we aim to preserve the privacy of the code while offering the ability to inspect it without compromising its secrecy, we propose an approach divided into two main phases -- \emph{Code Privacy} phase and \emph{Encrypted Code Analysis} phase.
Figure~\ref{fig:arch} shows \name's architecture, with the first phase running on the Developer's side (i.e., in its PC/infrastructure), and the second phase running on the Analyser's side.

The \emph{Code Privacy} phase (blue flow in the figure) prepares the code for encryption by extracting its DCFG, which captures the code's data and control flows, and representing it as an inverted index~\cite{manning2008introduction,curtmola2006searchable}, which will enable SSE queries on those flows, as a static data flow analysis would be carried out. Then, the index is encrypted following an SSE-like encryption scheme we designed, resulting in an \emph{Encrypted Inverted Index} (EII) that can be securely stored in an untrusted location. This phase is executed on the Developer's side (i.e., in its PC/infrastructure), and it corresponds to CCA's $\mathsf{Encrypt}$ protocol.

The \emph{Encrypted Code Analysis} phase (the orange and green flows in the figure) allows developers or authorised analysts to perform analysis tasks for vulnerability detection in the encrypted code, i.e., over the EII. The result of this phase is the vulnerable encrypted data flows, which are reported back to the developer for decryption and remediation of the vulnerabilities found. This phase corresponds to protocols $\mathsf{Authorise}$ and $\mathsf{Analyse}$, thus involving both the developer and analyser, but its heavier computations are executed on the analyser's side.

Nonetheless, the solution is transparent for both sides, i.e., developers only need to submit their code and respond to Authorise requests, and analysers just select a \emph{Code Analysis Task} (i.e., a vulnerability class to test), and \name's modules perform the required computations and output the results.
Sections~\ref{sec:ITL} and \ref{sec:Det_Vuln} will detail these two main phases, respectively.
Additionally, Figure~\ref{fig:arch_example} showcases \name processing an example PHP program (in the left top corner) for an XSS vulnerability detection task, including the output of each module. The program contains an XSS vulnerability flowing through lines \{1,8\}. 

\begin{figure*}[h]
	\centering
	\includegraphics[width=1\textwidth]{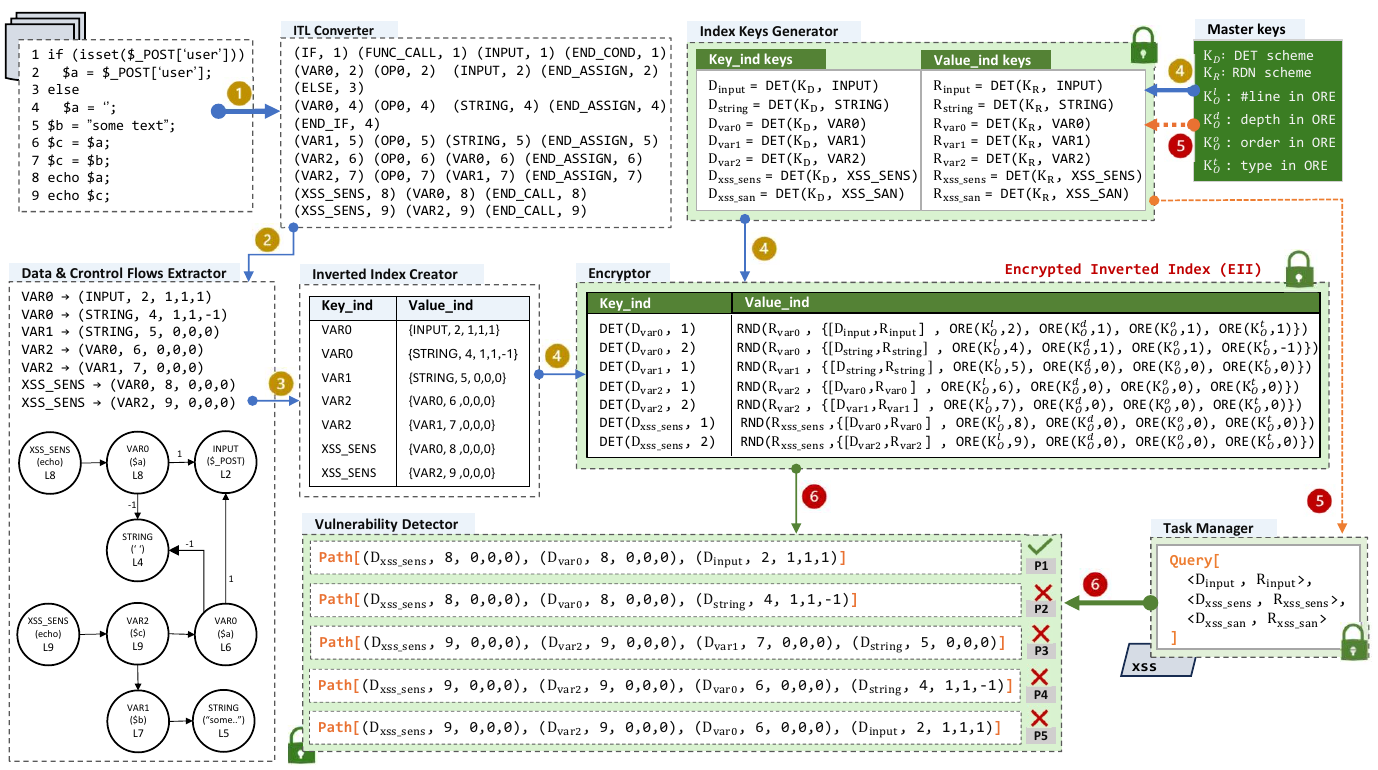}
	%\vspace{-8mm}
	\caption{Example of executing \name with the XSS vulnerability detection analysis task in PHP. The encryption of numerical values in the Vulnerability Detector box is omitted for simplicity.}
	\label{fig:arch_example}
	%\vspace{-3mm}
\end{figure*}

%%%%%%%%%%%%%%%%%%%%%%%%%%%%%%%
\section{Code Privacy Phase}
\label{sec:ITL}

This phase involves processing and preparing the application code through a five-stage pipeline that encrypts it while keeping data dependencies and control flow data. These stages are carried out by modules \emph{ITL Converter}, \emph{Data \& Control Flows Extractor}, \emph{Inverted Index Creator}, \emph{Index Keys Generator} and \emph{Encryptor}.

%----------------------
\subsection{Intermediary Token Language (ITL) Converter}

The application code is parsed and split according to the language specification (e.g., PHP) and its tokens, following, therefore, its syntactic and semantic aspects.
The result is a LexToken stream composed of LexTokens~\cite{AhoLSU2006,levine1992lex}, where each one represents a broken part of the code, denoted by a token defined in the language specification, along with the code it represents and its position in the code.
LexTokens are therefore expressed in the form of \texttt{<token\_type, value, line\_number>}.
The type of LexTokens (\texttt{token\_type}) can be, for example, \texttt{VAR} to represent variables, \texttt{FUNC\_CALL} for function calls, \texttt{IF} and \texttt{FOR} for keywords of the language, and \texttt{LPAREN} for the left-parentheses metacharacter.
Afterwards, LexTokens are normalised to remove irrelevant data (e.g., comments) and translated into an \emph{Intermediate Token Language} (ITL), thus resulting in an \emph{ITL-token} stream that preserves the logic, semantics, and data and control flows of the source code.
The ITL must be simple, lightweight, and generic enough to enable static analysis and cryptographic techniques over it, while also abstracting/obfuscating the names of variables, operators, and user functions.
To achieve this, the ITL Converter uses two databases -- \emph{ITL Tokens \& Rules} and \emph{Vuln. Knowledge} -- during the ITL translation process (explained in §~\ref{sec:translation}).

%-----------------
\subsubsection{ITL-Tokens \& Rules and Vuln. Knowledge Databases}
\label{sec:class_knowledge}
These databases contain rules and code elements that are needed to convert the target programming language to ITL, as well as elements that can be interesting to search given the intended analysis tasks.

The first database defines ITL-tokens and establishes how to process and translate LexTokens to them (detailed in §~\ref{sec:translation}). The tokens used in our ITL are mainly provided by the \texttt{token\_type} of LexTokens, plus those specified for code analysis tasks we want to support (e.g., those presented in Table~\ref{tab:tokens_example}) and others that we define for specific purposes, such as denoting the end of assignments, conditions, and control flow branches. In addition, an ITL-token can represent several code elements with the same meaning (e.g., \texttt{\$\_POST} and \texttt{\$\_GET} entry points are translated into a single \texttt{INPUT} ITL-token).

\begin{table}[t]\scriptsize\footnotesize
	%\vspace{-3mm}
	\caption{Example of a Vulnerability (Vuln.) Knowledge database for detection of XSS and SQLi vulnerabilities in PHP.}
	\label{tab:tokens_example}	
	\centering
	%\vspace{-3mm}
	%\addtolength{\tabcolsep}{-1.5}
	\begin{tabular}{|l|l|}
		\hline
		\textbf{ITL-Token} & \textbf{PHP code element}  \\
		\hline\hline
		INPUT          & \$\_GET, \$\_POST, \$\_FILES, \$\_SESSION, \$\_COOKIE \\ \hline
		XSS\_SENS      & echo, print, exit  \\ \hline
		SQLi\_SENS     & mysqli\_query, mysqli\_stmt\_execute, mysqli\_execute \\ \hline
		XSS\_SAN      & encodeForHTML, htmlentities, htmlspecialchars  \\ \hline
		SQLi\_SAN     & mysqli\_real\_escape\_string, mysqli\_stmt\_bind\_param   \\ \hline
	\end{tabular}
	%\vspace{-3mm}
\end{table}

The second database associates vulnerability-related ITL-tokens to their respective code elements from the target programming language. 
This can include entry points, sensitive functions, and sanitisation functions (functions that invalidate malicious inputs) which are then mapped to ITL-tokens related to specific vulnerability classes.
Table~\ref{tab:tokens_example} shows an example of this for the XSS and SQLi vulnerability classes in PHP.
For XSS, for instance, the functions \texttt{echo} and \texttt{htmlentities} are, respectively, a sensitive sink and a sanitisation function, which are mapped to \texttt{XSS\_SENS} and \texttt{XSS\_SAN} ITL-tokens.

%-------
\subsubsection{Translation process}
\label{sec:translation}

Each group of LexTokens belonging to the same line of code is translated into an ITL line, after removing LexTokens of metacharacters that are irrelevant for representation as ITL and analysis (e.g., the left-parenthesis, \texttt{LPAREN}). After cleaning, each remaining LexToken of the code line is analysed and transformed into a tuple \texttt{<ITL-token, line\_number>}, where the \texttt{ITL-token} will be either the LexToken's \texttt{token\_type} or a specific token for the code analysis task (e.g., XSS detection), and the \texttt{line\_number} is the line number presented in the LexToken. Afterwards, an \emph{ending} ITL-token is added to the resulting ITL line, i.e., one of those we created to this effect, such as \texttt{END\_ASSIGN} to denote the end of an assignment instruction.

\vspace{1mm}
\noindent
\textbf{Knowledge about specific analysis tasks.}
for each LexToken of the type \texttt{FUNC\_CALL} or \texttt{VAR}, the translator matches its \texttt{value} with the data in the Vuln. Knowledge database, replacing such tokens with ones specific for the vulnerability classes considered.
For instance, for PHP, the \texttt{VAR} LexToken is replaced by the \texttt{INPUT} ITL-token if its \texttt{value} starts with $\$\_$, thus indicating that it is an entry point. 
Also, the \texttt{FUNC\_CALL} LexToken is replaced by an ITL-token from Table~\ref{tab:tokens_example} if its \texttt{value} matches any function from there. For example, the \texttt{XSS\_SENS} ITL-token is chosen if the LexToken's value matches the sensitive function \texttt{echo}.

\vspace{1mm}
\noindent
\textbf{Names of variables, user functions and operators.}
concrete token names for variables, user functions and operators are replaced by abstract ITL-tokens since, for analysis tasks, we only need to be able to distinguish different ITL-tokens, regardless of their concrete names. Such an abstraction also adds an extra layer of obfuscation and privacy. Thus, the token \texttt{VAR} (representing variables) is adapted to be different for each variable by appending an incremental counter to the end of the ITL-token. 
For example, in the code of Figure \ref{fig:arch_example}, both variables \texttt{\$a} and \texttt{\$b} are initially represented by the LexToken \texttt{VAR}; the translator determines that the ITL-token for the former will be \texttt{VAR0}, since \texttt{\$a} appears first (line 2), while for the latter it will be \texttt{VAR1}. As the variable \texttt{\$a} reappears later in the code (lines 4, 6 and 8), it will again be replaced by \texttt{VAR0} because the translator stores the context to track the variables that have already appeared.
For user function names, the same procedure is applied, using the ITL-token \texttt{FUNC} followed by an incremental counter to distinguish the different user functions. 
Also, for operators, the translator follows the same process but with a slight difference: tokens representing different operators (e.g., \texttt{PLUS} for $+$, \texttt{EQUALS} for $=$) are replaced by a single ITL-token -- \texttt{OP} -- followed by an incremental counter, resulting in operators such as \texttt{OP0, OP1, OP2}.

The ITL Converter box of Figure \ref{fig:arch_example} depicts the result of this translation process. 
We can observe that abstract ITL-tokens \texttt{VAR0} and \texttt{OP0} represent, respectively, the variable and operator, \texttt{INPUT} represents an entry point, and \texttt{END\_ASSIGN} the \emph{ending} of the instruction. Note that \texttt{FUNC\_CALL} also appears, which denotes built-in functions that are not relevant for vulnerability detection purposes but are retained for semantic effects.

%---------------------
\subsection{Data \& Control Flows Extractor}

Finding code interconnectivity over the ITL-token stream means finding the stream's existing data and control flows. While the former relates to data dependencies between ITL-tokens, the latter is associated with conditional ITL instructions. This module is responsible for obtaining both types of flows. While processing ITL-tokens, on the one hand, it creates connections of data dependencies and data flows throughout the code by analysing assignments and function calls, and, on the other hand, it extracts control flows by analysing conditional instructions. The resulting flows can be expressed as a graph, where the nodes are ITL-tokens, extended with the information needed to represent branched control flows, and the edges are connections. We call this graph the \emph{Data and Control Flow Graph} (DCFG).

%-----
\subsubsection{Data Dependency and Bottom-up Data Flow}
\label{sec:data_dep}
Typically, data dependency is formed by connections, in which variables are used in assignment instructions and predicates throughout the program~\cite{yamaguchi,ferrante}. We have extended this concept by also considering as connections variables that appear as parameters of function calls, since task analysis can also involve these cases. 

As \name's \emph{Encrypted Code Analysis} phase relies on the reasoning behind SSE, i.e., an \emph{inverted} index~\cite{curtmola2006searchable}, the Data \& Control Flows Extractor module must follow an inverted flow approach to build the data dependency graph, and thereby enable subsequent \emph{bottom-up} data flow analysis.
Therefore, data dependencies within assignment instructions will flow from left to right, e.g., for the \texttt{\$a = \$\_POST['user'];} instruction, portrayed in ITL as \texttt{(VAR0,2)(OP0,2)(INPUT,2)(END\_ASSIGN, 2)}, the module creates a connection pointing from \texttt{VAR0} to \texttt{INPUT}, i.e., \texttt{VAR0 -> INPUT} (meaning that \texttt{\$a} points to \texttt{\$\_POST['user']}).
We recall that \name keeps the context of variables and operators that appear throughout the stream, and, therefore, the module can determine when an \texttt{OP$_x$} ITL-token is an assignment operator in order to establish the connections correctly.
In the case of function calls, data dependencies flow from the function name to its arguments, e.g., \texttt{XSS\_SENS -> VAR0}.
In Figure \ref{fig:arch_example}, at the bottom of the Data \& Control Flows Extractor box, the DFCG is shown for the program we have been following. Observing the graph, we can see, for example, the dependencies of \texttt{VAR0} (\texttt{\$a} variable) with lines 2, 4, 6 and 8.

%-----
\subsubsection{Control Flow}
\label{sec:controlFlow}

Since static analysis does not evaluate the code while it is being executed, it cannot determine which branch of the program's control flow to follow. So, it needs to consider every branch that can occur, i.e., understand the program's control flow of conditional instructions (\texttt{if}, \texttt{switch} and \texttt{loops}) and whether or not one branch is independent of another. 
\texttt{switch} instructions are processed as being \texttt{if}, while \texttt{loops} are considered to be executed once and, hence, their instructions are processed as being outside the loops.

To this end, we extended the ITL tuples with three fields: \texttt{depth}, \texttt{order} and \texttt{type}, resulting in tuples in the form \texttt{<ITL-token, line\_number, depth, order, type>}. To help explain the purpose of these fields, we present in Listing \ref{lst:cfnumbers} a PHP excerpt with various control flow branches, where these fields are illustrated as comments.
\texttt{depth} indicates the number of \texttt{if} statements nested inside another.
For instance, the depth of ITL-tokens of line 3 is 2 because this line exists within a nested \texttt{if} statement of depth 1. Line 8 also has depth 2 for the same reason and is not nested inside the \texttt{if} of line 2.
\texttt{order} denotes the number of \texttt{if} statements inside a depth.
For instance, the order of any token in line 3 is 1 because it belongs to the first if-else block inside the \texttt{if} of line 1.
Line 8 has order 2 since it belongs to the second \texttt{if} from the \texttt{if} of line 1.
Finally, \texttt{type} is the type of the control flow, and its value describes whether the branch is an \texttt{if} (value 1), an \texttt{elseif} (a positive number that increments for each \texttt{elseif}), or an \texttt{else} (value -1).
The type of line 3 is 1 because it is inside an \texttt{if}, but in line 5, it is -1 because it is inside an \texttt{else}.
Yet, these fields are set to zero in line 12 since they are outside any control flow if-branch.

\begin{figure}[t]
	\setcounter{figure}{0}
	\renewcommand{\figurename}{Listing}
	\begin{lstlisting}%
		[language=PHP,numbers=left,showstringspaces=false,xleftmargin=0.5cm, captionpos=b]
if(1==1){
   if(1==1)
      $a = $_POST['u']; //[depth=2,order=1,type=1]
   else
      $a = $_POST['u']; //[depth=2,order=1,type=-1]
	
   if(1==1)
      echo $a;         //[depth=2,order=2,type=1]
}
else
   $a = $_GET['u'];     //[depth=1,order=1,type=-1]
echo $a;                //[depth=0,order=0,type=0]
	\end{lstlisting}
	%\vspace{-3mm}
	\caption{A PHP program with various control flow branches and with different depths, orders and types, and their values.}
	\label{lst:cfnumbers}
	%\vspace{-5mm}
\end{figure}

Combining the two approaches described above, the Data \& Control Flows Extractor uses the ITL conditional instructions to fill these three fields and builds the DCFG by linking a series of pairs. 
In Figure \ref{fig:arch_example}, at the top of the Data \& Control Flows Extractor box, these pairs are illustrated. For each pair, the left part indicates the ITL-tokens with dependencies, while the right part shows their ITL tuples dependencies, represented in their extended form. 
For example, the module for the second ITL instruction creates a connection pointing from \texttt{VAR0} to \texttt{INPUT}, i.e., the pair \texttt{VAR0 -> (INPUT, 2, 1,1,1)}, meaning that \texttt{\$a} points to \texttt{\$\_POST['user']}, and that instruction is within a true \texttt{if}-branch. We can also observe that the second pair is the ITL instruction within the \texttt{else}-branch, and all the other instructions that do not present control have their control flow fields set to 0. Note that ITL conditional statements are used only to extract control flow and thus to fill these three fields. So, they do not need to be represented in DCFG.

%---------------------
\subsection{Inverted Index Creator}
\label{sec:InvInd}

An \emph{Inverted Index}~\cite{manning2008introduction} is a data structure that represents data in the form of \texttt{<key\_ind,value\_ind>} entries, where keywords (stored in \texttt{key\_ind}) are mapped to the documents in which they appear in (stored in \texttt{value\_ind}), with \texttt{key\_ind} entries being unique so as to avoid ambiguities between keywords and the documents they map.
An example application is unique keywords pointing to the IDs of text documents that contain them.
The index is queried for keywords and returns the documents in which they appear. Each time a new keyword appears in a document, a new entry is added to the index, representing that occurrence and linking the keyword to the document.

After constructing the DCFG, the inverted index is built from there. Each DCFG pair will be an entry in the index, where the ITL-token (the left side of the pair) will be the \texttt{key\_ind} and the extended ITL tuple (the right side of the pair) will be the index's \texttt{value\_ind}. In the corresponding box of this module in Figure~\ref{fig:arch_example}, we can observe the resulting inverted index from the DCFG.

%---------------------
\subsection{Encryptor}
\label{sec:encrypt}
Given the inverted index, the Encryptor module builds an encrypted version of it -- the \emph{Encrypted Inverted Index} (EII) --  that will support code analysis tasks.
This is achieved by adapting cryptographic techniques traditionally used for searching encrypted databases and text documents.  We begin this section with a brief overview of these tecnhiques, then explain the approach used by the Encryptor module.
%followed by an explanation of Encryptor.

\subsubsection{Searchable Symmetric Encryption (SSE)}
SSE~\cite{praticalsearchable} is a cryptographic technique that allows a client to securely store and query a database in an untrusted remote server. In common among most approaches is the dependence on an inverted index~\cite{curtmola2006searchable},  thus allowing queries to be performed in time sub-linear with the database size. 

The index is usually encrypted using a combination of two encryption schemes, a deterministic (\emph{DET}) and a probabilistic/randomised (\emph{RND}).
 In a \emph{DET} scheme, encrypting a message several times with the same cryptographic key will always produce the same ciphertext. Such a scheme is usually not considered secure since it leaks repetitions, unless the plaintext domain is unique (i.e., it holds no repetitions). Nonetheless, it supports an interesting feature: \emph{equality-testing} in the encrypted domain, i.e., testing whether two DET encryptions are equal. 
 In practice, there are many ways to build DET schemes, such as using a secure block cipher with fixed IVs or using a Hash-based Message Authentication Code (HMAC) (even though HMACs do not hold a decryption function, they can still be used to implement DET if decryption is not necessary).
 In contrast, an \emph{RND} scheme will always produce different ciphertexts, even when encrypting a message multiple times with the same cryptographic key. This more secure scheme can be used to protect data, but it does not allow any computation in the encrypted domain. \emph{RND} can be implemented by using a block cipher with random IVs, such as AES-CBC~\cite{modern_crypto}.

In SSE, \texttt{key\_ind} is encrypted with \emph{DET} while \texttt{value\_ind} is encrypted with \emph{RND}. Thus, the resulting EII index entries take the form \texttt{<DET($\text{K}_\text{D}$,key\_ind), RND($\text{K}_\text{R}$,value\_ind)>}, with \texttt{$\text{K}_\text{D}$} and \texttt{$\text{K}_\text{R}$} being cryptographic keys for \emph{DET} and \emph{RND} respectively.
This allows leveraging the equality-testing feature of \emph{DET} to search in the encrypted index while its encryptions reveal nothing, since by definition of inverted indexes, each \texttt{key\_ind} is unique. As such, information leakage can only occur when a query is performed.

\subsubsection{Order Revealing Encryption (ORE)}
ORE~\cite{lewi2016order} is a type of encryption scheme that preserves order relations between plaintexts, thus allowing order comparisons of numerical values to be performed after encryption. In ORE, a numerical value and a cryptographic key are taken as input, and a random byte string is produced as a result of encryption. Then, given two ORE encryptions, a comparison function outputs which encrypted value is higher, lower, or if equal, as if their plaintexts values were being compared. ORE is typically used in database systems that need to support range queries. In \name we use ORE to encrypt line numbers and numerical control flow information (namely \texttt{depth}, \texttt{order}, and \texttt{type}) of the index's \texttt{value\_ind}, so that these can be compared during code analysis without exposing their plaintext values.

%\vspace{-1mm}
\subsubsection{Building the EII for Encrypted Code Analysis}
\label{sec:se_code_analysis}

To build the Encrypted Inverted Index (EII), \name\ uses an approach similar to SSE, augmented with ORE so that no index data is ever exposed in plaintext, as described next:

\begin{itemize}
	\item The entries of the inverted index will be represented as SSE index entries (i.e., EII entries). The unique nature of \texttt{key\_ind} will be used to represent ITL-tokens, whereas \texttt{value\_ind} will store the extended ITL tuples to which ITL-tokens are connected to;
    \item For each distinct ITL-token found in the inverted index, two cryptographic keys (one for \emph{DET} and one for \emph{RND}) will be generated. 
    These keys will be used to encrypt \texttt{key\_ind} and \texttt{value\_ind}, respectively, and to replace ITL-tokens in the extended ITL tuples.
    \item The information stored in \texttt{value\_ind} will allow traversing the EII, by linking to the next node of the DCFG graph (i.e., an EII entry) and storing the cryptographic keys needed to decrypt it.  
    \item Numerical data of the \texttt{value\_ind} (i.e., line numbers and control flow information) will be encrypted with \emph{ORE} so that it can be used during code analysis without exposing its plaintext values;
    \item Each type of numerical data in the EII will be encrypted with a different ORE key, since we do not need to compare order relations between different types (e.g., line numbers are only compared between each other, and not with \texttt{depth} information).
\end{itemize}

The obtained EII can then be traversed as if it was the DCFG in its regular form, with index entries being decrypted as they are accessed but without ever exposing any plaintext data from the source code.
Next, we explain our solution in detail.

\vspace{1mm}
\noindent
\textbf{Cryptographic keys generation and usage.}
\name, through the \textit{Index Keys Generator} module, employs six master keys: $K_D$ and $K_R$ to generate keys for the \emph{DET} and \emph{RND} schemes respectively, and $K_O^l,K_O^d,K_O^o,K_O^t$ to respectively encrypt line numbers, \texttt{depth}, \texttt{order} and \texttt{type} information with the ORE scheme. 

Then, for each distinct ITL-token found in the DCFG, two additional cryptographic keys are generated by encrypting it with the \emph{DET} scheme and keys $K_D$ and $K_R$, respectively. 
For instance, in the Index Keys Generator box of Figure~\ref{fig:arch_example}, two cryptographic keys for the ITL-token \texttt{VAR0} are generated by encrypting it with \emph{DET} and master keys $K_D$ and $K_R$, thus producing \texttt{$\text{D}_\text{var0}$=DET($\text{K}_\text{D}$,VAR0)} and \texttt{$\text{R}_\text{var0}$=DET($\text{K}_\text{R}$,VAR0)}. 
These keys are then used to encrypt index entries related to \texttt{VAR0}, namely \texttt{$\text{D}_\text{var0}$} is used with \emph{DET} to encrypt \texttt{VAR0}'s \texttt{key\_ind} and \texttt{$\text{R}_\text{var0}$} with \emph{RND} to encrypt \texttt{value\_ind} connected to \texttt{VAR0}'s \texttt{key\_ind}.
In addition, they take the place of the ITL-token in the extended ITL tuple of \texttt{value\_ind}, thus ensuring complete data secrecy while enabling EII traversal.

\vspace{1mm}
\noindent
\textbf{Composition of \texttt{key\_ind} and ensuring its uniqueness.}
Since we can have multiple assignments to the same variable and function calls of the same function in different code lines, we need a way to store this information in the index while guaranteeing that: $i)$ each \texttt{key\_ind} is still unique and $ii)$ the index does not leak the frequency of each assignment and function call.
Similarly to~\cite{cash2014}, we add a counter to each \texttt{key\_ind} specific to the ITL-token, i.e., each token will have a unique counter. These counters start with value $1$ and are incremented each time a specific ITL-token is found in the inverted index (i.e., on the left-side of the DCFG's pairs). The EII's \texttt{key\_ind} is then calculated by encrypting the counter with the \emph{DET} scheme and the cryptographic key associated with the ITL-token --- \texttt{DET($\text{D}_\text{{t}}$,$\text{c}_\text{{t}}$)}, with \texttt{$\text{D}_\text{{t}}$} and \texttt{$\text{c}_\text{{t}}$} respectively being the cryptographic key and counter of the ITL-token.
For instance, in Figure~\ref{fig:arch_example}, in the Inverted Index Creator box, the \texttt{VAR0} token appears twice as \texttt{key\_ind}, meaning that in the EII there are two \texttt{key\_ind} entries using $D_{var0}$ with \emph{DET} (the first two rows of EII in the Encryptor box), one with the counter value of $1$ (\texttt{DET($\text{D}_\text{{var0}}$,1)}) and another with the counter value of $2$ (\texttt{DET($\text{D}_\text{{var0}}$,2)}). The same procedure is applied to \texttt{VAR2} and \texttt{XSS\_SENS} tokens for the same reason.
This approach guarantees that: all \texttt{key\_ind} depend on their respective ITL-tokens; they are always unique; and they ensure the privacy of the ITL-token since it is never exposed in plaintext.

\vspace{1mm}
\noindent
\textbf{Composition of \texttt{value\_ind} and ensuring EII traverse.} 
We need to build \texttt{value\_ind} in a way that allows us to traverse and decrypt the EII as we perform data flow analysis tasks, but only entries relevant to a task should be decryptable and the information from the source code should never be exposed in plaintext. 
Additionally, and in contrast to traditional SSE, here we are not querying for keywords, but for data dependencies and control flows. Hence, we need to not only create entries for each ITL-token, but also establish the respective connections between them.

Inspired from~\cite{curtmola2006searchable}, we use \texttt{value\_ind} to point to the next EII entry from the data flow and store the cryptographic keys needed to access and decrypt it.
More specifically, for each index entry, we replace the ITL-token of the extended ITL tuple with its corresponding cryptographic keys and encrypt the whole tuple with the \emph{RND} scheme.
Example: consider the index entry \texttt{<XSS\_SENS, \{VAR0,8,0,0,0\}>} from Figure~\ref{fig:arch_example} (line 6 of the Inverted Index Creator box) and its encryption as presented in the Encryptor box. Its EII \texttt{value\_ind} contains $D_{var0}$ and $R_{var0}$, which are the cryptographic keys needed to access and decrypt the next entries of the data flow (i.e., lines 1 and 2 in the same box). These cryptographic keys represent the ITL-token \texttt{VAR0}, replacing it in the EII extended ITL tuple of line 6. Then, the whole tuple is encrypted with \emph{RND} and key $R_{xss\_sens}$, thus obtaining \texttt{RND($\text{R}_\text{{xss\_sens}}$, \{$\text{D}_\text{{var0}}$,$\text{R}_\text{{var0}}$, \dots\})}.

\vspace{1mm}
\noindent
\textbf{Keeping numeric values secret in \texttt{value\_ind}.}
To ensure full code privacy and no information leakage from decrypting \texttt{value\_ind} during code analysis, our approach also encrypts the line number and control flow information contained in each extended ITL tuple (i.e., \texttt{depth}, \texttt{order} and \texttt{type} fields). Since this data is only used in order comparisons, the four fields are encrypted using \emph{ORE}.
Additionally, since these fields do not need to be compared with each other, each can be encrypted using a different master key -- $K_O^l,K_O^d,K_O^o,K_O^t$ --, as described previously. 
In the index entry  \texttt{<VAR0, \{INPUT,2,1,1,1\}>} of our example, the four numeric values of \texttt{INPUT} will be encrypted as follows:
\texttt{(ORE($\text{K}_\text{O}^\text{l}$,2),ORE($\text{K}_\text{O}^\text{d}$,1),ORE($\text{K}_\text{O}^\text{o}$,1),ORE($\text{K}_\text{O}^\text{t}$,1))}.

\vspace{1mm}
\noindent
\textbf{The final form of EII entries.}
The next formula summarises the transformation from a \texttt{<}$\text{DCFG}_\text{t}$, $\text{DCFG}_\text{t+1}$\texttt{>} pair to an EII entry of the form \texttt{<key\_ind, value\_ind>}, where $t$ is an ITL-token, $t+1$ is the next token in the stream (i.e., $\text{DCFG}_\text{t}$ points to $\text{DCFG}_\text{t+1}$), and $c$ is a counter. Additionally, Figure~\ref{fig:inst_transf} illustrates the complete transformation of the instruction \texttt{\$a = \$\_POST['user']} (line 2 from Figure~\ref{fig:arch_example}) into an EII entry.
{\center
	\small

    \texttt{<$\text{DCFG}_\text{{t}}, \text{DCFG}_\text{{t+1}}$>} $=>$
    \texttt{<DET($\text{D}_\text{t}$,$\text{c}_\text{t}$), RND($\text{R}_\text{t}$,$\text{EII}_\text{{t+1}}$)>} :\\  
	%\vspace{0.1cm}
	\texttt{$\text{D}_\text{t}$ = DET($\text{K}_\text{D}$,$\text{t}$), 
		$\text{c}_\text{t}$ = \{$1$,...\},
		$\text{R}_\text{t}$ = DET($\text{K}_\text{R}$,$\text{t}$),}\\
	\texttt{$\text{EII}_\text{{t+1}}$ = \{$\text{D}_\text{{t+1}}$,$\text{R}_\text{{t+1}}$,ORE($\text{K}_\text{O}^\text{l}$,line\_n),ORE($\text{K}_\text{O}^\text{d}$,depth),\\
		\hspace{2cm} ORE($\text{K}_\text{O}^\text{o}$,order),ORE($\text{K}_\text{O}^\text{t}$,type)\}}
}

\begin{figure}[t]
	   %\vspace{-4mm} 
	\setcounter{figure}{2}
	\centering
	\includegraphics[width=1.02\columnwidth]{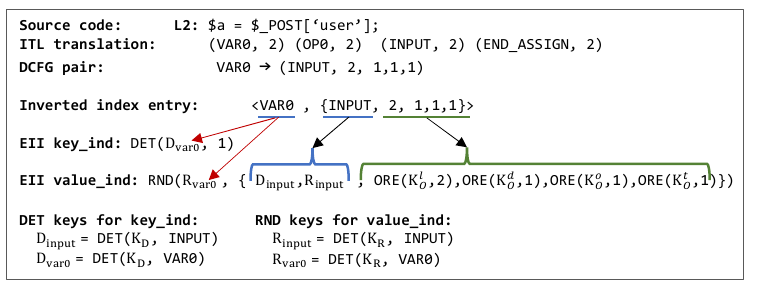}
	%\vspace{-8mm}
	\caption{Example of the transformation of an instruction of code into an EII entry.}
	\label{fig:inst_transf}
	%\vspace{-4mm} 
\end{figure}

%\input{sections/5-Code-Analysis}
%%%%%%%%%%%%%%%%%%%%%%%%%%%%%%%%
\section{Encrypted Code Analysis Phase}
\label{sec:Det_Vuln}

At the end of the code privacy phase (CCA protocol $\mathsf{Encrypt}$), the encrypted inverted index (EII) is obtained, and the developer can give the analyser access to it, i.e., to themselves or someone else from their software company, or even external software testers from third-party testing companies. 
To perform code analysis tasks (CCA protocol $\mathsf{Analyse}$), the analyser chooses the tasks it wants to perform and asks the developer to provide the cryptographic keys of the ITL-tokens (for \emph{DET} and \emph{RND} schemes) that are associated with those tasks (protocol $\mathsf{Authorise}$).
This is an important step, as it gives the developer an extra layer of access control over who can perform analysis tasks and which tasks they can perform. Given those keys, the analyser starts accessing the encrypted index, following the relevant data flows and decrypting index entries as it goes, and in the end gets the results of the tasks. In more detail, and as shown in Figure~\ref{fig:arch}, this phase is carried out by the collaboration of modules \emph{Task Manager} and \emph{Vulnerability Detector} on the analyser side, and \emph{Query Authoriser \& Creator} on the developer side. 

%\vspace{-1mm}
%--------------
\subsection{Task Manager}
This module receives an \emph{Analysis Task} chosen by the analyser, such as the XSS vulnerability class for finding this type of vulnerability in the developer's code. 
Then the module interacts with the developer's \emph{Query Authoriser \& Creator} module, to ask for the developer's authorisation and create the query subjacent to the analysis task. If authorised, the \emph{Vuln. Knowledge} database is accessed to retrieve the information previously defined for that task, namely the ITL-tokens associated with it (i.e., the ones representing entry points, sensitive and sanitisation functions of that vulnerability class), and then the cryptographic keys of the selected ITL-tokens are generated (through the Index Keys Generator module), thus composing the query with these pairs of keys.
For instance, for the \emph{XSS} analysis task, the ITL-tokens would be \texttt{INPUT}, \texttt{XSS\_SENS}, and \texttt{XSS\_SAN}, and the resulting query would be \texttt{Query[<$\text{D}_\text{{input}}$,$\text{R}_\text{{input}}$>, <$\text{D}_\text{{xss\_sens}}$,$\text{R}_\text{{xss\_sens}}$>, <$\text{D}_\text{{xss\_san}}$,$\text{R}_\text{{xss\_san}}$>]}.
Lastly, the module sends the query to the \emph{Vulnerability Detector} module to proceed with the analysis task, outputting its final result to the analyser. 
Formally, the steps taken to produce an authorised query implement CCA protocol $\mathsf{Authorise}$.

%\vspace{-1mm}
%----------------
\subsection{Vulnerability Detector}
This module implements protocol $\mathsf{Analyse}$. Specifically, it 
takes the cryptographic keys for querying the index, extracts the code paths for the \emph{Analysis Task}, and checks them for the vulnerability class, following the next four-step algorithm.

\vspace{1mm}
\noindent
\textbf{\emph{1) Finding data flow paths (DFP).}}
This step starts by accessing the EII index with the cryptographic keys, following a bottom-up data flow approach. We recall that due to the nature of the inverted index, in which function call flows must be represented in an inverted way, and so their names appear as \texttt{key\_ind}, we must follow a bottom-up approach. This means that we start the analysis by the end-point (last flow) of a vulnerability search -- i.e., a sensitive sink. Hence, in our example, we start the analysis with $D_{xss\_sens}$ and $R_{xss\_sens}$ keys, as they mark the last flow of an XSS vulnerability.
However, remember that due to possibly repeated assignments and function calls, \texttt{key\_ind} is composed not only of the encrypted token $D_{t}$ but also of its counter $c_{t}$. This means that to start accessing the index, the Vulnerability Detector must combine $D_{t}$ and $c_{t}$ through the \emph{DET} scheme, as was previously done by the Encryptor to compose \texttt{key\_ind} and ensure \texttt{key\_ind} uniqueness. 
Moreover, as neither the vulnerability detector module nor the analyser has any knowledge about the EII index content, and thereby does not know how often a given token is repeated, the module will start with $c_{t}$=1 and check if the resulting ciphertext (i.e., \texttt{DET($D_{t}$,1)}) exists as a EII's \texttt{key\_ind}, continuing with increasing counter values until one fails to be found. For instance, in the example of Figure~\ref{fig:arch_example}, \texttt{$D_{xss\_sens}$} appears twice (in the Encryptor box), meaning that the module will create and test \texttt{DET($D_{xss\_sens}$,1)}, \texttt{DET($D_{xss\_sens}$,2)}, and \texttt{DET($D_{xss\_sens}$,3)}, stopping with the last one as it will not be found in the index. 
For each \texttt{key\_ind} entry that is found, its corresponding \texttt{value\_ind} is decrypted with $R_{t}$, and the resulting data flows are followed until their end is reached (i.e., a terminal ITL-token like \texttt{INPUT}), repeating the same procedure. 
The result is a list of encrypted data flow paths (DFP) that will be processed by the next steps of the algorithm to determine whether they are vulnerable. Each DFP starts with an ITL-token from a \texttt{key\_ind}, followed by one or more extended ITL tuples provided by the corresponding \texttt{value\_ind}.
Figure~\ref{fig:arch_example} shows this list with five DFP paths ($P1$ to $P5$), in the Vulnerability Detector box. 
For a better understanding and clarity of the tuples contained in the encrypted DFPs, we present their numerical data without the ORE encryption. But we recall that these DFPs are managed by the module and presented to the analyser completely encrypted, revealing nothing about their content.

\vspace{1mm}
\noindent
\textbf{\emph{2) Removing invalid DFPs.}}
Next, the algorithm filters out invalid paths, such as those with tokens whose \texttt{line\_number} is greater than the \texttt{line\_number} of the DFP's first token but belong to the same scope (e.g., code block, user function). For example, suppose that on the path $P3$, the token \texttt{($D_{string}$,5,0,0,0)} had its \texttt{line\_number} equal to 12, which is greater than the \texttt{line\_number} of the first token \texttt{($D_{xss\_sens}$,9,0,0,0)}. $P3$ would be an invalid path because line 12 cannot be executed before line 9.
Hence, these kinds of DFPs can be safely discarded, as they clearly do not exist in the context of the program.

\vspace{1mm}
\noindent
\textbf{\emph{3) Aggregating DFPs.}} 
As we follow a bottom-up approach, the starting DFP's token is the \texttt{final\_token} associated with the code analysis task. So, this step groups and sorts DFPs by this final token and its line number, i.e., \texttt{<final\_token, line\_number>}, thus creating sub-list of such pairs. For the $D_{xss\_sens}$ final token, in our example, the sub-lists would be [P1, P2] for \texttt{<$D_{xss\_sens}$,8>} and [P3, P4, P5] for \texttt{<$D_{xss\_sens}$,9>}.

\vspace{1mm}
\noindent
\textbf{\emph{4) Outputting the final DFPs.}}
This step determines which paths are correct, considering the program's control flow and whenever \texttt{depth} is different from zero. %, control flow branches exist. 
It employs three analyses over each sub-list of DFPs resulting from step 3): $(i)$ resolving control flow branches, $(ii)$ obtaining the closest DFP to the \texttt{final\_token}, and $(iii)$ determining whether a path is vulnerable. 

In $(i)$, for each distinct \texttt{depth}, paths are separated into sets: one with all paths with different \texttt{orders} regardless of \texttt{type}, and others including paths with equal \texttt{order} but different \texttt{type} (one set for each \texttt{order}). The first set will contain paths in which variables are propagated between \texttt{if}s, while the other sets will contain paths of the same \texttt{if}.
Based on Listing \ref{lst:cfnumbers} and the sink of line 8, examples of the first case are the paths \texttt{(2,2,1), (2,1,-1)} (lines 8 and 5) and \texttt{(2,2,1), (2,1,1)} (lines 8 and 3). Examples of the second case are, in our example, the paths [P1, P2] and [P4, P5]. Therefore, after applying this step, our initial sub-lists are organised as follows [P1,P2] and [P3, [P4,P5]].

In $(ii)$, for the resulting sets of a \texttt{depth} and for the paths whose tokens have these three fields equal, excluding the \texttt{final\_token} or also excluding the last token of the path, the algorithm finds the closest path to the final token.
In other words, it looks for the first difference in the paths and checks which one is the closest to the \texttt{final\_token}.
%if P2's control flow equals P1's, i.e., \texttt{(1,1,1)},
For example, supposing that P1 and P2 have equal control flow, including their last token, i.e., \texttt{(1,1,1)}, both paths would have these three fields equal, and the first difference between them would be the \texttt{line\_number} of their last token, hence P2 would be chosen because its \texttt{line\_number} is the one closest to \texttt{($D_{xss\_sens}$,8,0,0,0)}. 
In our example, as this scenario is not observable in both sub-lists from $(i)$, they follow as such to step $(iii)$.

In $(iii)$, the resulting paths are checked for the vulnerability class subjacent to the analysis task, relying on three determinations:
$(a)$ for the resulting DFPs of $(ii)$, all paths ending with \texttt{$D_{input}$} token are chosen and checked which ones do not contain a sanitisation token (e.g., \texttt{$D_{xss\_san}$ for XSS});
$(b)$ for the resulting sub-sets of DFPs of $(i)$ that did not fall in $(ii)$ and in which the \texttt{type} field of their last tuple differs, the paths ending with \texttt{$D_{input}$} token are chosen and checked which ones do not contain a sanitisation token. 
In our example, sub-sets [P1,P2] and [P4,P5] fall into this scenario. P2 and P4 are discarded, as their last tuple does not end with \texttt{$D_{input}$}. So, from [P1,P2] and [P3,[P4,P5]] sub-lists came from $(i)$, results [P1] and [P3,P5]; 
$(c)$ for the resulting paths of $(a)$ and $(b)$ within sub-lists of the \texttt{final\_token}, the step $(ii)$ is applied, followed by the filter $(a)$. 

Following our example, at this point, the sub-list from \texttt{<$D_{xss\_sens}$,8>} is [P1] and the sub-list from \texttt{<$D_{xss\_sens}$,9>} is [P3, P5]. 
Applying step $(ii)$ and filter $(a)$ to this second sub-list results in no paths, meaning therefore that both paths are dropped. Despite P3 results from $(ii)$ (thus dropping P5), it is discarded by $(a)$ because it does not end with \texttt{$D_{input}$} token. 
The sub-list containing only [P1] is checked for the presence of the \texttt{$D_{input}$} token as its last token and for the absence of sanitisation tokens. $P1$ satisfies both conditions, therefore considered vulnerable, and thus it is the final result of verification $(iii)$ and of the whole vulnerability detection process.

Lastly, the vulnerable DFPs are outputted by the Vulnerability Detector, which returns them to the Task Manager and subsequently to the analyser. The analyser can then share the results with the developer and provide suggestions on improving the source code based on the vulnerability class it chooses to analyse. 
In our example of Figure~\ref{fig:arch_example}, only $P1$ is outputted as an XSS vulnerability. 
In this case, the analyser may suggest using the sanitisation function \texttt{htmlentities} to remove vulnerabilities of XSS.

%\input{sections/6-Security}
%%%%%%%%%%%%%%%%%%%%%%%%%%%
\section{Security Analysis}
\label{sec:security}

Our goal with \name is to securely implement the functionality idealised for \emph{Confidential Code Analysis} (CCA). This means that: $(i)$ its index should not reveal anything about the code; $(ii)$ and its query tokens, when combined with the index, should only reveal the result of the intended analysis task. Formally, we will use the leakage function $\mathcal{L^{\name}}=(\mathcal{L}^{\name}_{Encrypt},\mathcal{L}^{\name}_{Authorise},\mathcal{L}^{\name}_{Analyse})$ to capture this information and analyse \name's security.

Demonstrating $(i)$, we have an encrypted index (the EII) composed of multiple $<$\texttt{key\_ind}, \texttt{value\_ind}$>$ entries. \texttt{value\_ind} encryptions will not reveal anything since they are produced by \emph{RND} and all have the same size before and after encryption. \texttt{key\_ind} encryptions also do not reveal anything, since even though encrypted with \emph{DET}, each represents either a different ITL-token or the same token but different counter value. Hence, the only thing leaked by the EII is its size, which represents the total number of DCFG's pairs.
We deem this as basic information, nonetheless it could also be hidden through padding with dummy index entries. This means that $\mathcal{L}^{\name}_{Encrypt} = \bot$.

Demonstrating $(ii)$, we must first analyse the leakage of query tokens by themselves and then when combined with EII. For the first part, query tokens are encrypted with \emph{DET}, which leaks repetitions. This means that when invoking the $\mathsf{Authorise}$ protocol, the adversary learns the deterministic encrypted query tokens associated with a task of its choice. 
Formally, this must be modelled as $\mathcal{L}^{\name}_{Authorise} = \{\{D_0,R_0\},\dots,\{D_k,\dots,R_k\}\}$, where $\{D_i,R_i\}$ is an encrypted token pair associated with the query.

For the second part of $(ii)$, we must analyse the knowledge gained by the adversary when accessing the index with an encrypted query token. Indeed when an analysis task is performed, the adversary learns $\{P_0,...,P_n\}$, the set of paths accessed when solving the query, with $P_i=<T_0,...,T_n>$ being a path with multiple index entries $T_j=$\texttt{<$D_j$,$R_j$,$ORE^l_j$,$ORE^d_j$,$ORE^o_j$,$ORE^t_j$>} that were accessed, where $\{D_j,R_j\}$ is the \emph{DET} encrypted token pair associated with the entry, $ORE^l_j$ is the \emph{ORE} encryption of its line number, $ORE^d_j$ of its \texttt{depth}, and so on. Formally, this is modelled as $\mathcal{L}^{\name}_{Analyse} = \{P_0,...,P_n\}$.

Although $\mathcal{L}^{\name}_{Analyse}$ seems to be leaking many different things, this is either basic information or is part of the intended computation and, hence, unavoidable. $\{P_0,...,P_n\}$ represents the found paths.
We deem this as part of the result of the intended analysis task. Moreover, the adversary only learns the number of paths, their sizes, the \emph{DET} encrypted tokens in each, and the corresponding \emph{ORE} encryptions.
\emph{DET} encrypted tokens reveal co-occurrence of ITL-tokens between paths but are still encrypted and even if their encryptions would be broken, ITL-tokens have been obfuscated before encryption by replacing operators and variable/function names with abstract tokens. \emph{ORE} encrypted line number and control flow information reveal frequency and order relations, but only between each field (i.e., the adversary can compare $ORE^l_j$ with $ORE^l_i$ but not with $ORE^d_j$) since each is encrypted with a different key.

%\input{sections/7-Implementation}
%%%%%%%%%%%%%%%%%%%%%%%%%%%%%%%%
\section{\name Implementation \& Deployment}
\label{sec:Implment}

\subsection{Implementation}
We implemented a prototype of \name~\cite{CoCoA:23} in Python to process PHP applications for XSS and SQLi vulnerability detection tasks. 
All modules were developed from scratch, with the exception of the Lexer sub-module, which was based on PLY's Lex module~\cite{ply}.
Our ITL totals 40 distinct ITL-tokens: 23 from LexTokens, 12 for \emph{ending} instructions (e.g., assignments, conditionals), and 5 for XSS and SQLi vulnerabilities. 
Our current prototype covers almost the whole PHP specification, except for some less commonly used instructions (e.g., tri-dimensional or greater arrays) and some object-oriented instructions. 
Also, although it supports analysing user function calls, it still needs improvements at the inter-procedural level regarding files. 
To implement the cryptographic operations, we used SHA1-HMAC for the \emph{DET} scheme and AES-128 in CBC mode for \emph{RND}, through the \emph{Hashlib}, \emph{Hmac}, and \emph{PyCryptodome} Python libraries. For \emph{ORE}, we used the ORE scheme by Lewi and Wu~\cite{lewi2016order} and their \emph{FastORE} library implementation~\cite{fastore}, with 32-bit input size and 8-bit blocks.
Lastly, databases were built with YAML~\cite{yaml}.
\name can be extended to other code analysis tasks and programming languages -- Section~\ref{app:extensions} gives an overview of how this can be done.

%===============
\subsection{Deployment Scenarios}
\begin{figure}[t]
\centering
	\includegraphics[width=1\columnwidth]{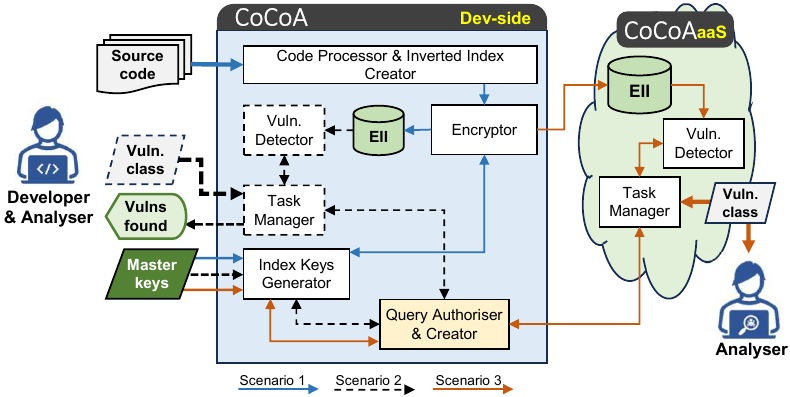}
	%\vspace{-6mm}
	\caption{Deployment scenarios of \name.}
	\label{fig:scenarios}
	%\vspace{-6mm}
\end{figure}

\name's approach was designed to ensure complete code privacy while enabling the analysis of the encrypted code by authorised software analysts. As previously stated, we envision three scenarios in which \name can be deployed and used (as illustrated in Figure~\ref{fig:scenarios}):

\vspace{1mm}
\noindent
\textbf{Scenario 1.} 
%\texttt{Scenario 1:} 
The tool is deployed on the developer's side, at the development software company, and is used only to protect the code and the developer's intellectual property (IP), i.e., encrypted code analysis is disabled. Even though this scenario is possible, it does not explore the full capacities of \name;

\vspace{1mm}
\noindent
\textbf{Scenario 2.} 
%\texttt{Scenario 2:} 
It is an extension of Scenario 1, in which \name is totally deployed and used on the developer side, allowing code analysis tasks over EII by analysers that could be developers themselves or the security team (e.g., purple team) existing in the software company. 

\vspace{1mm}
\noindent
\textbf{Scenario 3.} 
%\texttt{Scenario 3:}
\name is used as a service ({\name}aaS) by software testing companies, in which software companies contract the testing software service to external companies and authorise them to analyse their code in a protected manner. The deployment is bi-parted between the developer (software company) and the analyser (the testing company). {\name}aaS is deployed locally within the infrastructure's testing company or in the cloud, both comprising the EII index and the modules Task Manager and Vulnerability Detector, thus allowing code analysis tasks by the analyser. On the other hand, the remaining \name modules are deployed on the developer side, giving them control over their code and the tasks that can be performed over EII. This scenario is particularly interesting because, besides preserving the privacy of the code, protecting the developer's IP and allowing code analysis tasks by external analysts in a controlled manner, it also protects the testing companies against possible defamations of non-disclosure agreements (NDA) regarding the code, and allows them to perform the tasks remotely, i.e., out-of-house, instead of the standard in-house (of the software company). 

%\input{sections/8-Evaluation}
%%%%%%%%%%%%%%%%%%%%%%%%%%%%%%
\section{Evaluation}
\label{sec:impl_eval}

We divide our evaluation section between theoretical analysis and experimental evaluation.

%======
\subsection{Theoretical Analysis}
In this section, we perform a complexity analysis of \name regarding both communication and storage overheads. 
In \name, we have three protocols: $\mathsf{Encrypt}$, $\mathsf{Authorise}$, and $\mathsf{Analyse}$. As such, we evaluate each protocol separately.

Regarding $\mathsf{Encrypt}$, this protocol is executed solely on the developer's side. Nonetheless, the resulting EII can be shared with the analyser when the protocol finishes. As such, we consider a communication complexity of $O(n)$, where $n$ is the size of the EII in terms of the number of data flows and dependencies extracted from the source code. As for storage complexity, the developer has to store six master keys of constant size (O(1)), and either it or the analyser stores the EII. Hence, the storage complexity is $O(1+n)=O(n)$.

For the $\mathsf{Authorise}$ protocol, the analyser sends an analysis task to the developer (which we can assume to be of constant size, e.g., ``XSS'') and in return receives the cryptographic keys of the associated ITL-tokens if authorised, or nothing if not. Hence, the communication complexity of this step is $O(t)$, where $t$ is the number of ITL-tokens associated with the task. In terms of storage complexity, we assume $0$ (zero) since the exchanged data is temporary, only needed to perform the task, and can be discarded later.

Finally, for the $\mathsf{Analyse}$ protocol, this step is performed mostly by the analyser, but the final results can be shared with the developer. As such, we consider a communication complexity of $O(r)$, where $r$ is the size of the result set of the analysis task (e.g., the number of vulnerable paths for the XSS analysis task). In terms of storage, we again consider a complexity of $0$ (zero), since the resulting set is temporary and can be discarded after the developer fixes its code.

Concluding, \name's overall communication complexity is $O(n+t+r)=O(n)$ (since $t$ and $r$ must forcibly be a subset of $n$) and its storage complexity is $O(n)$.

%======
\subsection{Experimental Evaluation}

The objective of the experimental evaluation was to answer the following questions:
$i)$ can \name perform code analysis with high precision and a low false positive rate?
$ii)$ how do \name's results fare compared to standard (non-confidential) static analysis tools?
$iii)$ what are the performance and storage overheads imposed by our approach?

%============
\subsubsection{Vulnerability Detection Effectiveness} 
\label{sec:valvuln}

This section evaluates \name's effectiveness in detecting XSS and SQLi vulnerabilities. For this, we built and used two PHP application datasets -- one synthetic and one real-world -- and compared their results with 3 standard, non-confidential static analysis tools: Pixy~\cite{pixy}, PHPCorrector~\cite{Morgado:20}, and WAP~\cite{wap}. The datasets were made public and are included with \name's source code~\cite{CoCoA:23}. Additionally, the metrics used to evaluate the tools are defined in Table~\ref{tab:metrics}.

\begin{table}[t] %\footnotesize\scriptsize
    \caption{Metrics used to evaluate the tools in vulnerability detection}
    \label{tab:metrics}
    \centering
    %\vspace{-3mm}
    \resizebox{\columnwidth}{!}{%    
    \begin{tabular}{l|l}
    \hline
     \textbf{Metric} & \textbf{Definition and Formula} \\ \hline\hline
      Accuracy \textit{(acc)} & Measures how correctly the tool classifies instances as vulnerable \\ 
      & (TP - True Positive) and not-vulnerable (TN - True Negative)\\
      & over the total number of instances. \\ 
      & $acc=(\mathit{TP}+\mathit{TN})/(\mathit{TP}+\mathit{FN}+\mathit{TN}+\mathit{FP})$ \\
      Recall \textit{(rec)} & Measures the proportion of actual vulnerabilities that are \\
      & correctly identified by the tool. $rec=\mathit{TP}/(\mathit{TP}+\mathit{FN})$ \\
      Precision \textit{(pr)} & Measures the proportion of correctly identified vulnerabilities \\
      & out of all instances flagged as vulnerabilities by the solution.\\
      & $pr=\mathit{TP}/(\mathit{TP}+\mathit{FP})$ \\
      F1-score  & Measures the harmonic mean between \textit{pr} and \textit{rec}, providing \\
      & a single metric that balances the trade-off between the two. \\
      & $F1=2*(pr*rec)/(pr+rec)$ \\
      FP rate \textit{(fpr)} & Proportion of not-vulnerable (FP- False Positive) cases incorrectly \\
      & classified as vulnerable. $fpr=\mathit{FP}/(\mathit{FP}+\mathit{TN})$ \\
      FN rate \textit{(fnr)} & Proportion of vulnerable cases missed (FN - False Negative)\\
      & by the tool. $fnr=\mathit{FN}/(\mathit{FN}+\mathit{TP})$ \\ \hline
    \end{tabular}
    }
%\vspace{-5mm}
\end{table}

%------
\vspace{1mm}
\noindent
\textbf{Detection with Synthetic Applications.}
We built a synthetic ground-truth (GT) dataset comprising 3,205 SQLi and XSS PHP code excerpts provided by SARD~\cite{samate}. This GT includes a diversity of vulnerable (Vul) and not-vulnerable (N-Vul) cases with varying coding forms. These range from the standard use of sanitisation functions (e.g., \texttt{mysqli\_real\_escape\_string}) and appearance of vulnerabilities in the code to uncommon forms of validating user inputs (e.g., \texttt{filter\_var} and casting to data types). GT also includes object-oriented code and alternative methods for reading input data (e.g., \texttt{fgets}) besides the superglobal variables (e.g., \texttt{\$\_GET}, \texttt{\$\_POST}).
The dataset is balanced with 1,569 Vul and 1,636 N-Vul excerpts, containing 1,441 SQLi code excerpts (861 Vul and 580 N-Vul) and 1,764 XSS code excerpts (708 Vul and 1,056 N-Vul).

We executed \name and the 3 tools with this dataset.
Both \name and WAP processed all cases without errors. Pixy and PHPCorrector had trouble with 477 and 148 cases, respectively, due to the handling of object-oriented code and conditional instructions involving function calls.
Table~\ref{tab:gt} shows the results, expressed using the metrics in Table~\ref{tab:metrics}, calculated from the confusion matrices. Results are presented for each task in separate (XSS, SQLi) and combined (GT).

\begin{table}[b]\footnotesize\scriptsize
	%\vspace{-4mm}
	\caption{Evaluation \name and competitors with the GT dataset.}
	\label{tab:gt}
	%\vspace{-3mm}
	\centering
	%\resizebox{\columnwidth}{!}{
		\addtolength{\tabcolsep}{-1.6mm}
		\begin{tabular}{l|ccc|ccc|ccc|ccc}
			%\hline
			\cline{2-13}
			& \multicolumn{3}{c}{\textbf{\name\,}} & \multicolumn{3}{|c}{\textbf{Pixy}} & \multicolumn{3}{|c}{\textbf{PHPCorrector}} & \multicolumn{3}{|c}{\textbf{WAP}}  \\ \cline{1-13}
			\textbf{Metric} & \textbf{XSS}    & \textbf{SQLi}    & \textbf{GT} & \textbf{XSS}    & \textbf{SQLi}    & \textbf{GT} & \textbf{XSS}    & \textbf{SQLi}    & \textbf{GT} & \textbf{XSS}    & \textbf{SQLi}    & \textbf{GT} \\
			\hline\hline
			Precision (pr) & 0.70 & 0.97 & 0.80 & 0.54 & 0.74 & 0.64 & 1.00 & 0.73 & 0.86 & 0.35 & 0.65 & 0.39 \\
			Recall (rec)   & 0.30  & 0.19 & 0.24 & 0.61 & 0.87 & 0.74 & 0.02 & 0.02 & 0.02 & 0.50 & 0.12 & 0.29 \\
			Accuracy (acc) & 0.67 & 0.51 & 0.60 & 0.64 & 0.71 & 0.66 & 0.61 & 0.42 & 0.53 & 0.42 & 0.43 & 0.43 \\
			F1-score       & 0.42 & 0.32 & 0.37  & 0.57 & 0.80 & 0.68 & 0.04 & 0.03 & 0.03 & 0.41 & 0.20 & 0.33 \\
			FP rate (fpr)  & 0.10 & 0.00 & 0.06 & 0.35 & 0.60 & 0.41 & 0.00 & 0.01 & 0.01 & 0.63 & 0.09 & 0.44 \\
			FN rate (fnr)  & 0.70  & 0.81 & 0.76 & 0.39 & 0.13 & 0.27 & 0.98 & 0.99 & 0.98 & 0.50 & 0.89 & 0.71 \\
			\hline
		\end{tabular}
		%}
	%\vspace{-3mm}
\end{table}

\name presented similar precision (\emph{pr}) to that of PHPCorrector, and both outperformed Pixy and WAP. PHPCorrector achieved the highest \emph{pr} but also the lowest recall (\emph{rec}). Pixy achieved the highest \emph{rec}, but also showed a higher false positive rate (\emph{fpr}), which can be counterproductive for this type of tools. In contrast, \name's \emph{fpr} was much lower, and its \emph{rec} and F1-score aligned with WAP. We believe these lower results are not due to \name's design but to limitations in our current prototype implementation, especially regarding inter-procedural analysis and object-oriented instructions, and thus can still be improved.

%------
%Table da secao Det vuln with real apps
\begin{table*}[t]\footnotesize\scriptsize
%\vspace{-4mm}
\caption{Results and metrics of \name, Pixy, PHPCorrector and WAP with real web applications.}
\label{tab:vulnfind}
%\vspace{-3mm}
\centering
%\resizebox{\columnwidth}{!}{
	\addtolength{\tabcolsep}{-1.35mm}
	\begin{tabular}{l|cccc|cccc|cccc|cccc|l|cccc}
		%\hline
		\cline{2-17}
		\multicolumn{1}{c}{} & \multicolumn{4}{|c}{\textbf{\name\,}} & \multicolumn{4}{|c}{\textbf{Pixy}} & \multicolumn{4}{|c}{\textbf{PHPCorrector}} & \multicolumn{4}{|c}{\textbf{WAP}} & \multicolumn{5}{|c}{} \\ \cline{1-22}
		\textbf{WebApp} & \textbf{XSS}    & \textbf{SQLi}    & \textbf{FP}   & \textbf{FN}  & \textbf{XSS} & \textbf{SQLi} & \textbf{FP} & \textbf{FN} & \textbf{XSS} & \textbf{SQLi} & \textbf{FP} & \textbf{FN} & \textbf{XSS} & \textbf{SQLi} & \textbf{FP} & \textbf{FN} & \textbf{Metric} & \textbf{\name\,} & \textbf{Pixy} & \textbf{PHPCorrector} & \textbf{WAP} \\
		\hline\hline
		CurrentCost~\cite{currentcost} & 0 & 3 & 0 & 5 & 5 & 3 & 3 & 0 & 0 & 0 & 0 &  8 & 4 & 3 & 2 & 1 & \textbf{pr} & 0.93 & 0.72 & 0.83 & 0.93 \\
		DVWA~\cite{dvwa}  & 13 & 7 & 2 & 0 & 0 & 4 & 2 & 16 & 2 & 0 & 0 & 18 & 2 & 4 & 2 & 14 & \textbf{rec} & 0.56 & 0.63 & 0.36 & 0.64 \\
		Ghost~\cite{ghost} & 8 & 1 & 0 & 1 & -- & -- & -- & -- & 5 & 1 & 0 & 4 & 5 & 2 & 0 & 3 & \textbf{acc} & 0.61 & 0.56 & 0.42 & 0.67 \\
		Peruggia~\cite{peruggia} & 2 & 1 & 0 & 17 & -- & -- & -- & -- & 5 & 10 & 2 & 5 & 3 & 15 & 0 & 2 & \textbf{F1-score} & 0.70 & 0.67 & 0.50 & 0.76 \\
		Samate~\cite{samate} & 15 & 3 & 2 & 1 & 11 & 4 & 1 & 4 & 4 & 0 & 0 & 15 & 11 & 3 & 0 & 5 & \textbf{fpr} & 0.18 & 0.64 & 0.32 & 0.23 \\
		WackoPicko~\cite{wackopicko} & 2 & 0 & 0 & 9 & -- & -- & -- & -- & 8 & 0 & 5 & 3 & 5 & 3 & 0 & 3 & \textbf{fnr} & 0.44 & 0.37 & 0.64 & 0.36 \\
		Zipec~\cite{zipec} & 0 & 0 & 0 & 10 & 7 & 2 & 8 & 1 & 0 & 0 & 0 & 10 & 0 & 3 & 1 & 7 &&&&&\\
		\hline\hline
		%\hhline{=================}\cline{18-22}
		%\cline{1-22}\cline{1-17}
		\textbf{Total}           & \textbf{40}     & \textbf{15}      & \textbf{4}  &  \textbf{43}   & \textbf{23}  & \textbf{13}   & \textbf{14} & \textbf{21} & \textbf{24} & \textbf{11} & \textbf{7} & \textbf{63} & \textbf{30}  & \textbf{33}   & \textbf{5} & \textbf{35} &\multicolumn{5}{|c}{}\\
		\cline{1-17}
	\end{tabular}
	%}
%\vspace{-4mm}
\end{table*}

%---------
\vspace{1mm}
\noindent
\textbf{Detection with Real Applications.}
\label{sec:evalvuln}
We further assessed \name's effectiveness by analysing seven open-source PHP real web applications with known vulnerabilities, selected from~\cite{wap}, and comparing its results with the 3 tools. 
Table~\ref{tab:vulnfind} shows the results obtained. The XSS and SQLi columns represent the number of vulnerabilities correctly found by the tools, while FP and FN columns show the number of false positives and false negatives, respectively. 
The last five columns of the table show the metrics calculated from the confusion matrix of the tools, expressing, thus, further analysis of these results. 

\name had the best \emph{pr} -- 93\% --, like WAP, and outperformed the other tools in the \emph{fpr} -- 18\% --, where it only wrongly detected 4 data flow paths as vulnerable. WAP was the best tool, but it yielded results similar to \name. 
\name and WAP correctly detected a similar number of vulnerabilities: 55 and 66; hence, \name had less \emph{rec} than WAP (56\% and 64\% respectively), meaning it missed a few more vulnerabilities. 
The \emph{F1-score} is 70\% for \name versus 76\% for WAP.
Overall, PHPCorrector exhibited the worst results in almost all metrics. 
It only outperformed Pixy in \emph{pr} (83\% and 72\% respectively).
Pixy was only capable of processing 4 out of the 7 applications, so only these were considered for the metrics. 

We also tested \name without encryption (i.e., by disabling the Encryptor module). The results were the same as with encryption and hence omitted for simplicity. This shows that our encryption approach does not affect analysis effectiveness.

Comparing the results obtained with the two datasets and taking the \emph{F1-score} metric as a reference, as it considers $pr$ and $rec$ and thus gives a good idea of a tool's effectiveness, \name kept its second-best tool but doubled its \emph{F1-score} (70\%), probably justified by the absence of unusual coding forms and object-oriented code in some of these real applications. 
Contrarily, when we analyse its $fpr$, it increased from 6\% to 18\%, which means that it wrongly classified more data flow paths as vulnerable. This was due to inter-procedural analysis between files not being as precise in our current prototype.
We recall that each instance of GT is a single file, whereas real applications comprise more than one file. So, this increment in $fpr$ was expected.

%============
\subsubsection{Performance Overhead}  
\label{sec:evaltime}

%%%%%%%
\begin{table*}[t]\footnotesize\scriptsize%
\caption{Execution time in milliseconds (ms) of the \name modules, and comparison with WAP, for the XSS task, including encryption overheads.}
\label{tab:lexTrans}
%\vspace{-3mm}
\centering
%\resizebox{\textwidth}{!}{
	\addtolength{\tabcolsep}{-1.6mm}
	\begin{tabular}{l|cc|c|c|c|ccc|ccccc|ccc|ccccc||c}
		%\hline
		\cline{4-23}
		\multicolumn{3}{c|}{} & \textbf{Prep} & \textbf{~Lexer~} & \textbf{Trans} & \multicolumn{3}{c|}{\textbf{DCF and Encryptor}} & \multicolumn{5}{c|}{\textbf{Code Privacy Phase Total}} & \multicolumn{3}{c|}{\textbf{Vuln. Detector}} & \multicolumn{5}{c||}{\textbf{Total Execution (CP+VD)}} & \textbf{WAP}\\ \hline %\cline{4-6}
		& & & \textbf{Exec} & \textbf{Exec} & \textbf{Exec} & \textbf{DCF} & \textbf{DCF} & \textbf{DCF} & \textbf{CP} & \textbf{CP} & \textbf{Enc} & \textbf{CP} & \textbf{ORE} & \textbf{VD} & \textbf{VD} & \textbf{VD} & \textbf{Total} & \textbf{Total} & \textbf{Enc} & \textbf{Total} & \textbf{ORE} & \textbf{Exec}\\ %\cline{1-3}
		\textbf{WebApp} & \textbf{F} & \textbf{LoC} & \textbf{time} & \textbf{time} & \textbf{time} & \textbf{-Enc} & \textbf{+Enc} & \textbf{+ORE} & \textbf{-Enc} & \textbf{+Enc} & \textbf{OH (\%)} & \textbf{+ORE} & \textbf{OH (\%)} & \textbf{-Enc} & \textbf{+Enc} & \textbf{+ORE} & \textbf{-Enc} & \textbf{+Enc} & \textbf{OH (\%)} & \textbf{+ORE} & \textbf{OH (\%)} & \textbf{time}\\ 
		
		\hline
		\hline
		CurrentCost	&	2	&	60	  &	1.21	&	0.47	&	1.23	&	0.38	&	1.44	&	9.78	&	3.3	    &	4.35	&	32.16	&	12.69	&	285.15	&	0.02	&	0.04	&	0.04	&	3.31	&	4.39	&	32.57	&	12.73	&	284.26	&	113	\\
		DVWA	    &	8	&	291	  &	1.50	&	2.37	&	5.52	&	1.55	&	5.98	&	45.08	&	10.94	&	15.37	&	40.52	&	54.47	&	397.95	&	1.39	&	4.1	    &	18.33	&	12.33	&	19.48	&	57.9	&	72.81	&	490.3	&	153	\\
		Ghost	    &	14	&	460	  &	6.24	&	5.67	&	7.1	    &	2.13	&	8.66	&	68.12	&	21.13	&	27.66	&	30.88	&	87.13	&	312.26	&	2.26	&	6.74	&	38.96	&	23.39	&	34.4	&	47.06	&	126.08	&	439.06	&	163	\\
		Peruggia    &	6	&	589	  &	3.49	&	3.54	&	4.95	&	2.12	&	7.77	&	62.55	&	14.09	&	19.74	&	40.1	&	74.52	&	428.79	&	0.97	&	2.93	&	18.02	&	15.06	&	22.68	&	50.57	&	92.54	&	514.5	&	217	\\
		Samate	    &	7	&	164	  &	1.18	&	1.46	&	4.66	&	1.25	&	4.91	&	36.02	&	8.55	&	12.21	&	42.75	&	43.32	&	406.59	&	0.89	&	3,00	&	8.53	&	9.44	&	15.21	&	61.13	&	51.85	&	449.25	&	153	\\
		WackoPicko  &	36	&	2,099 &	17.43	&	14.25	&	25.74	&	9.42	&	32.88	&	263.35	&	66.84	&	90.29	&	35.09	&	320.77	&	379.92	&	0.45	&	1.26	&	1.84	&	67.29	&	91.56	&	36.06	&	322.61	&	379.44	&	301	\\
		Zipec	    &	7	&	588	  &	4.16	&	04.06	&	5.16	&	2.27	&	7.27	&	53.83	&	15.66	&	20.65	&	31.87	&	67.22	&	329.18	&	0.31	&	1.12	&	2.92	&	15.98	&	21.78	&	36.31	&	70.15	&	339.04	&	169	\\
		\hline\hline
		
		\textbf{Total}	&	\textbf{80}	&	\textbf{4,251}	&	\textbf{35.22}	&	\textbf{31.83}	&	\textbf{54.35}	&	\textbf{19.12}	&	\textbf{68.89}	&	\textbf{538.73}	&	\textbf{140.51}	&	\textbf{190.29}	&	\colorbox{light-gray}{\textbf{35.42}}	&	\textbf{660.13}	&	\colorbox{light-gray}{\textbf{369.79}}	&	\textbf{6.29}	&	\textbf{19.2}	&	\textbf{88.64}	&	\textbf{146.8}	&	\textbf{209.48}	&	\colorbox{light-gray}{\textbf{42.7}}	&	\textbf{748.77}	&	\colorbox{light-gray}{\textbf{410.05}}	&	\textbf{1,269}	\\
		\hline
	\end{tabular}
	%}
%\vspace{-4mm}
\end{table*}

To evaluate performance, we used the real dataset and conducted a series of tests with the XSS task to measure the execution time of each of \name's modules. Each test was repeated and measured five times, and their average values were obtained and presented in Table \ref{tab:lexTrans}.
For the tests, we used a laptop with an AMD Ryzen 5 5600H CPU at 3.3GHz, 16Gb of RAM, and 512GB SSD, running Manjaro Linux 23.1.

\vspace{1mm}
\noindent
\textbf{Preprocessor.} 
The preprocessor (Prep in the table) is a module that does not figure in \name's design, but was added to our prototype to filter non-PHP code and non-PHP files from the dataset.
Depending on the web application, this module can take from 1 ms to 17 ms to execute. After its execution, 94 PHP files (F) were ready for processing. However, only 80 were processed effectively by the tool, as the remaining 14 files contained object-oriented and advanced PHP instructions that the tool can not handle yet.

%------
\vspace{1mm}
\noindent
\textbf{ITL Converter.}
In \name's implementation, the ITL Converter module includes the Lexer and the ITL Translator, so we evaluate these components separately.
The Lexer took approximately 32 ms to process all 7 webapps, which consisted of over 4,200 lines of code (LoC) spread over 80 files. 
As expected, the execution time tends to increase with the number of LoC and files. More LoC implies more tokens and data flows to process, and more files implies more system calls, which is also time-consuming.
For the ITL Translator (Trans), some apps took approximately the same time to process as in the Lexer, while others took almost double the time. We believe this is due to the number of LoC and also the number of tokens per LoC in those applications, as this module needs to convert all LexTokens to ITL.

%------
\vspace{1mm}
\noindent
\textbf{Data \& Control Flows Extractor and Encryptor.}
We tested the Data \& Control Flows Extractor in three scenarios: $(i)$ with the Encryptor module disabled (DCF -Enc in the table); $(ii)$ with \emph{DET} and \emph{RND} encryption of the index (DCF +Enc); and $(iii)$ additionally with \emph{ORE} encryption of line numbers and control flow (DCF +ORE). We separated the evaluation of ORE as it increases overhead by a large fraction.

As expected, the Data \& Control Flows Extractor by itself adds little overhead to the solution. However, with \emph{DET} and \emph{RND} encryptions, this overhead grows by a large percentage, as many encryptions have to be done per index entry. Still, this overhead can be considered practical, as even in the worst case (WackoPicko webapp, with 36 files and 2,099 LoC) this module took only 32 ms to conclude. \emph{ORE} encryptions, however, can increase overhead by an order of magnitude in some cases. This is due to the complexity of the encryption algorithm and also the number of encryptions (four) per index entry.

The column \textit{Code Privacy Phase Total} in Table~\ref{tab:lexTrans} shows the sum of all steps described so far, with additional encryption (\emph{DET} and \emph{RND}) and \emph{ORE} overheads as a percentage, for the convenience of the reader.

%------
\vspace{1mm}
\noindent
\textbf{Vulnerability Detector.}
This module was tested with an XSS analysis task and similarly to the Data \& Control Flows Extractor: without encryption, with \emph{RND} and \emph{DET} encryptions, and additionally with \emph{ORE}. Looking at Table~\ref{tab:lexTrans}, we can verify, once again, that \emph{DET} and \emph{RND} encryptions cause execution time to increase by some fraction and \emph{ORE} by an order of magnitude in some cases. However, since this module is very fast, its overhead, even with \emph{ORE}, can still be considered practical, with 38 ms in the worst case (Ghost) and 88 ms total with all webapps.
Interestingly, we could also observe that this module's execution time is strictly related to the number of sensitive sinks (e.g., \texttt{XSS\_SENS}) present in the application.

%------
\vspace{1mm}
\noindent
\textbf{Total Execution Time.}
The total execution time of \name was also measured without encryption, with \emph{DET} and \emph{RND}, and additionally with \emph{ORE}, along with overhead percentages. 
Results show that the tool took a total of 146 ms to process all applications without encryption and 209 ms with \emph{DET} and \emph{RND}, i.e., an encryption overhead, on average, of 42.7\%.
With \emph{ORE}, total overhead increased to 410.05\%, totalling 748 ms to process all webapps and perform the XSS analysis.

Although these overheads may indicate some drawback of \name, it should be noted that a large part of its execution time comes from the code privacy phase (see columns 10-14 of Table~\ref{tab:lexTrans}), which typically will only be executed once per app. Thus, of the encryption overheads, 35.42\% and 369.79\% are associated with this phase, respectively. The remaining overhead comes from the encrypted code analysis phase (7.28\% and 40.26\%, respectively), which is low and practical for the benefit gained in privacy.

\vspace{1mm}
\noindent
\textbf{Comparison with WAP.}
The last column of Table~\ref{tab:lexTrans} shows the time that WAP took to process the same webapps.
%In comparison, 
\name revealed to be much more efficient than WAP; even with \emph{ORE} encryptions. However, we believe the main reason for this discrepancy is that WAP not only detects vulnerabilities, but also predicts whether those vulnerabilities are real. 
Nonetheless, we can observe that WAP's execution time increases with LoCs in a similar fashion as our solution.

%============
\subsubsection{Memory and Storage Overhead}
\label{sec:evalspace}

\begin{figure}[t]
\centering
\includegraphics[width=1\columnwidth]{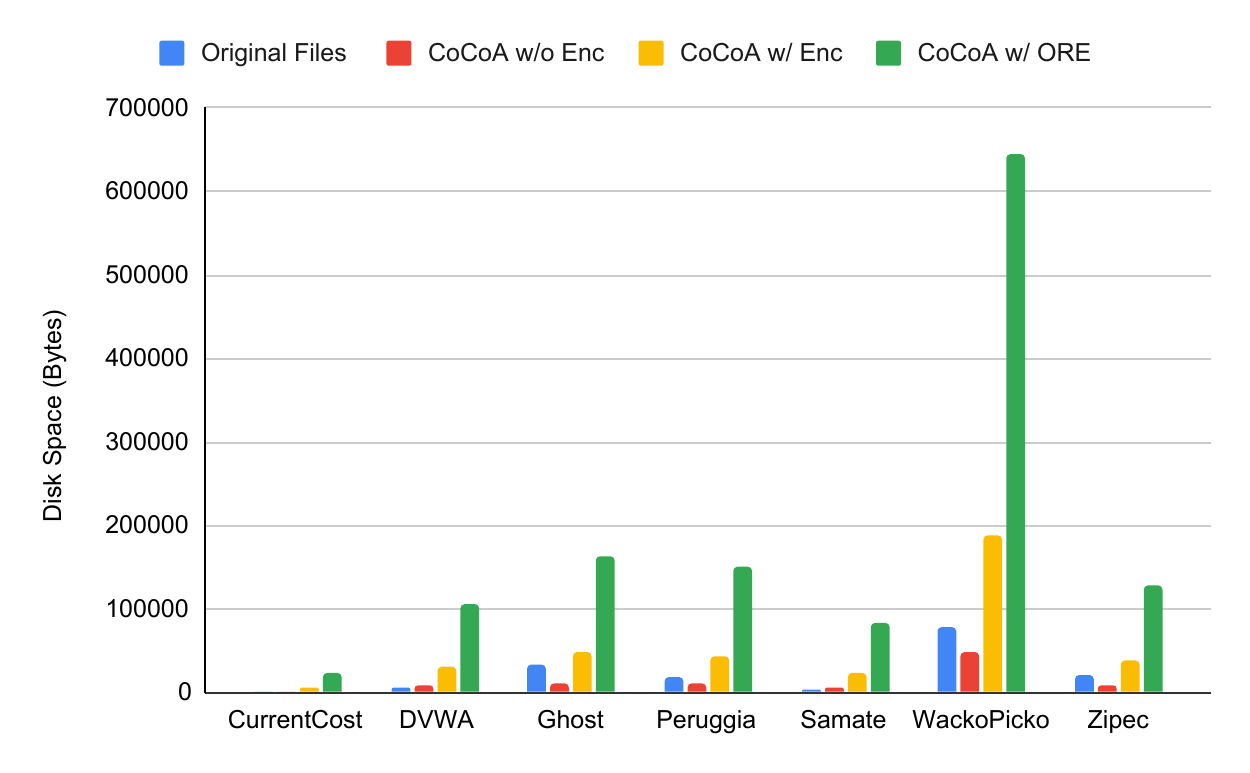}
%\vspace{-10mm}
\caption{Storage space used by the source code and the encrypted data structure, in bytes.}
\label{fig:mem}
%\vspace{-4mm}
\end{figure}

%As a last experiment, 
Lastly, we compared the storage space required by the source code of the applications with that of the index produced by \name without encryption, with \emph{DET} and \emph{RND}, and with \emph{ORE}.
Figure~\ref{fig:mem} illustrates the results of these scenarios.
\name's index without encryption occupied less space than the original webapps; this is to be expected since \name simplifies the code structure, discards irrelevant data, comments, and so on. Exceptions to this were the smallest applications with less than or equal to 7 KB, like Samate. Nonetheless, by adding \emph{DET} and \emph{RND} encryptions, the index size grew by some fraction due to the ciphertext expansion of these ciphers (e.g., 78 KB of source code to 49 KB of plaintext index and 188 KB of encrypted index in WackoPicko), and again by an order of magnitude with \emph{ORE} (645 KB in WackoPicko).

%%%%%%%%%%%%%%%%%%%%%%%%%%%%%%%%
\section{\name Extensibility}
\label{app:extensions}

In this section, we give an overview on how \name can be further extended to other analysis tasks and languages.

\vspace{1mm}
\noindent
\textbf{Analysis Tasks.}
Currently, \name's prototype performs two analysis tasks -- detection of SQLi and XSS vulnerabilities. It can be extended to other classes or even to a different type of analysis task. In both cases, 
%To extend to another class, 
the ITL-tokens that will represent the task and the code elements associated with that task should be added to the \emph{Vuln. Knowledge} database. For example, to include the detection of Remote File Inclusion (RFI) vulnerabilities, the ITL-token \texttt{RFI\_SENS} should be created, and the PHP code properties \texttt{include} (and others related to RFI) should be associated with this ITL token.
To accommodate a new type of analysis task, e.g. type checking related to integer overflows, ITL-Tokens should be created to denote the integer data type (e.g., \texttt{INT\_TYPE}), and arithmetic plus operations ($+$) should be considered throughout the data flow analysis to check if any exceeds the maximum value allowed for the data type.

\vspace{1mm}
\noindent
\textbf{Other Programming Languages.}
Extending \name to other programming languages requires a set of additions. Nonetheless, the existing ITL-tokens can be reused, as well as the DCFG data structure and its encryption method. More concretely, the \emph{ITL Tokens \& Rules} and \emph{Vuln. Knowledge} databases need to be constructed with the code properties and tasks of the new language.
For example, to extend \name to Java, we can reuse the \texttt{INPUT} ITL-token and map it to the \texttt{getParameter} entry point .
Other tokens, such as operators, reserved words (\texttt{IF}, \texttt{WHILE}), and variable identifiers, could be reused without changes.

%%%%%%%%%%%%%%%%%%%%%%%%%%%%%%
\section{Related Work}
\label{sec:RW}

\noindent\textbf{Static Analysis.}
Static analysis aims to find bugs in code without executing it \cite{staticviolations, principles}, by checking the code of a program~\cite{fie,wap,dekant,leopard}.
Usually, static analysers build a model of the code through lexers to break the code into tokens \cite{wap, rips}, parsers to generate code representation structures~\cite{ferrante, wap} that the analyser will traverse to execute a set of rules about something it intends to find and a knowledge base that contains the data that characterise what it will look for in the code (e.g., vulnerabilities). Data flow analysis is the most used method to traverse these structures, and taint analysis is the most common technique used to discover bugs and extract the associated data flows \cite{wap,leopard,last}.
Pixy \cite{pixy}, PHPCorrector \cite{Morgado:20}, RIPS \cite{rips}, and WAP \cite{wap} use this analysis to verify PHP code, efficiently detect sources and sinks, and analyse the data flow in the program through intra and inter-procedural techniques, resulting in accurate results.
These tools must access the source code in cleartext, unlike \name, and therefore do not provide code privacy.

%-----
\vspace{1mm}
\noindent\textbf{Code Obfuscation and Encryption.}
Code obfuscation and encryption are used to make code unreadable~\cite{Hosseinzadeh:18}. The reasoning behind this is to protect the source code from a possible copyright violation (e.g., a company reusing the code of another company) and reverse engineering analysis by making it illegible~\cite{Banescu:18,Kang:21,Dong:16,Cho:11}. There are several ways to obfuscate code \cite{yakproref,srcprotectorref,Hosseinzadeh:18,Wu:16}, including changing variable names, modifying the order in which the code is written, and relying on jumps to retain the correct execution order \cite{Banescu:18}. %Cappaert:08 %Heiderich:11
Although this technique seems effective in ensuring code privacy, it is reversible, as it does not use encryption to obfuscate the code.
In addition, this practice can make static analysis infeasible or less reliable since these tools are typically designed to parse the original source code without any form of obfuscation~\cite{Schrittwieser:16}. 
In contrast, code encryption has been used to ensure code privacy for cloud storage purposes~\cite{Dong:16} or to avoid reverse engineering analysis~\cite{Cho:11,Cappaert:06}, in which the entire or partial code (e.g., entry points and sensitive sinks) is encrypted and tested against tampering attacks. Code task analysis is allowed, but during the analysis process, the code is decrypted.  %Cappaert:08
\name aims to protect the code with encryption and, simultaneously, enable static analysis over encrypted data using SSE, without revealing the code. During the code analysis of \name, when decrypting the \texttt{value\_ind} of the EII, what is found are the cryptographic keys of the next EII entry and the numeric data encrypted with ORE; so nothing is revealed about the code. Furthermore, these keys are the representative form of ITL tokens, and ITL tokens, in turn, are a form of obfuscation of the actual source code.

%-----
\vspace{1mm}
\noindent\textbf{Confidential Computing.}
There are many encryption schemes that allow computations on encrypted data. Both Fully Homomorphic Encryption (FHE)~\cite{gentry2009fully} and Secure Multi-Party Computation (MPC)~\cite{cramer2015secure} support additions and multiplications on encrypted data and hence can be used to execute any computation in the encrypted domain. 
However, these computations must be represented as arithmetic/boolean circuits, which can be highly inefficient in some cases.
Moreover, FHE requires large CPU overhead and MPC large bandwidth overhead. Searchable Symmetric Encryption (SSE)~\cite{curtmola2006searchable,ferreira2013searching,ferreira2018muse} allows searching encrypted text databases efficiently, although revealing some information with each query. Oblivious RAM (ORAM)~\cite{stefanov2018path} eliminates this leakage, but at the cost of high bandwidth overhead.
SSE has also been used to encrypt graphs and perform shortest distance queries~\cite{meng2015grecs,ghosh2021efficient}. This is somehow similar to our approach; however, performing static analysis on a DCFG is different from performing a shortest-distance query on a graph, 
so an original approach had to be designed to solve our problem.

%%%%%%%%%%%%%%%%%%%%%%%%%%%%%%%%%%
%\vspace{-1mm}
\section{Conclusions}
\label{sec:conc}

The paper presented the initial study of \emph{Confidential Code Analysis}, a new research field that enables static code analysis tasks, such as vulnerability detection, in the encrypted domain while preserving code privacy. Also, \name is presented, the first tool in this field to detect vulnerabilities in PHP web applications. The tool combines static analysis and SSE to obtain an encrypted data structure in the form of an inverted index that represents the program's data and control flow in a way that allows searches on it, i.e., the execution of code analysis tasks as a data flow analysis considering control flow.
\name was tested on synthetic and real PHP web applications, and the results showed that it achieved 80\% and 93\% precision repectively, and overall better results than two other static analysis tools. The analysis tasks over encrypted code take no longer than on clear data, and the encryption overhead employed by the tool is minimal. 
Also, the proposed data structure to store the encrypted code requires less disk space than the source files.

%%%%%%%%%%%%%%%%%%%%%%%%%%%%%%%%%%
\section*{Acknowledgments}
This work was supported by P2030 through project I2DT, ref. COMPETE2030-FEDER-00389100, an ITEA4 European project (ref. 22025), and by FCT through the APOSTLE project (ref. \href{https://doi.org/10.54499/2023.12254.PEX}{2023.12254.PEX}) and the LASIGE Research Unit (ref. \href{https://doi.org/10.54499/UID/00408/2025}{UID/00408/2025}).

%\bibliographystyle{IEEEtranS}
%\bibliography{IEEEabrv,biblio}

%% FAZER no FIM
% bibliography section
\bibliographystyle{abbrv}

% Balance the references page
% \onecolumn
% \begin{multicols}{2}

%\vspace{-13mm}
%% FAZER no FIM
%% Sem foto
%\begin{IEEEbiographynophoto}
\begin{IEEEbiography}
[{\includegraphics[width=1.1in,height=1.25in,clip,keepaspectratio]{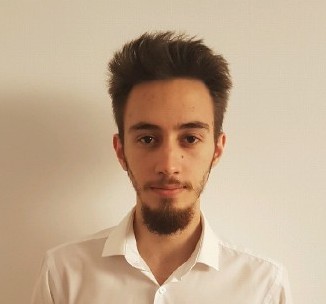}}]{Jorge Martins}
is a Software Engineer at Critical TechWorks - BMW Group - Portugal. He holds an MSc in Computer Engineering from the Faculty of Sciences of the University of Lisbon (FCUL). He is a young and passionate software engineer with an interest in Software Development Security. 
\end{IEEEbiography}

%\vspace{-15mm}
\begin{IEEEbiography}
[{\includegraphics[width=1.1in,height=1.25in,clip,keepaspectratio]{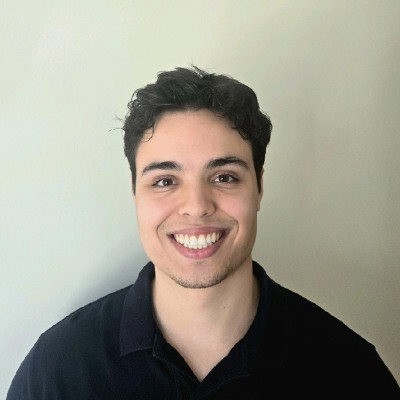}}]{David Dantas}
is a Software Engineer at ComplyAdvantage, Portugal. He holds an MSc in Computer Engineering from the Faculty of Sciences of the University of Lisbon (FCUL), where he also completed his BSc, and is a researcher of the LASIGE research unit. His thesis was related with cybersecurity, focusing on privacy-preserving systems, as part of the SMaRtChain project. His main interests include cybersecurity and distributed systems.
\end{IEEEbiography}

\vspace{-5mm}
\begin{IEEEbiography}
[{\includegraphics[width=1.1in,height=1.25in,clip,keepaspectratio]{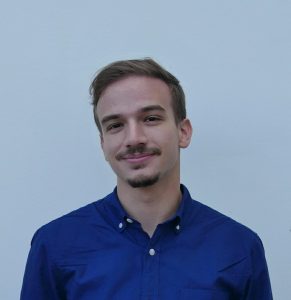}}]{Rafael Ramires}
is a PhD student of Informatics at the Faculty of Sciences of the University of Lisbon (FCUL) and the University of Luxembourg, doing research in the software security area. He holds an MSc in Computer Engineering from FCUL. He has been involved in activities for the detection of vulnerabilities in web applications in the SEAL, XIVT and I2DT projects.
His research interests include software security, artificial intelligence and multi-agent systems. 
\end{IEEEbiography}

%\vspace{-5mm}
\begin{IEEEbiography}
[{\includegraphics[width=1.1in,height=1.25in,clip,keepaspectratio]{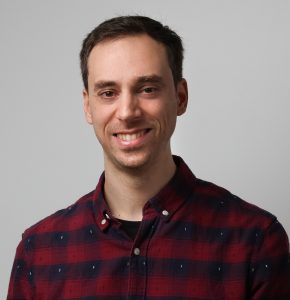}}]{Bernardo Ferreira}
is an Assistant Professor with the Faculty of Sciences, University of Lisbon, Portugal, and an integrated
researcher with the LASIGE research unit. He received the BSc, MSc, and PhD degrees in computer science from the Faculty of Sciences and Technology, Nova University of Lisbon, Portugal, in 2008, 2010, and 2016, respectively. His research interests include distributed systems security and privacy, with special focus on blockchains and secure distributed computation (Homepage: \url{https://www.di.fc.ul.pt/~blferreira/}).
\end{IEEEbiography}

%\vspace{-5mm}
\begin{IEEEbiography}
[{\includegraphics[width=1.1in,height=1.25in,clip,keepaspectratio]{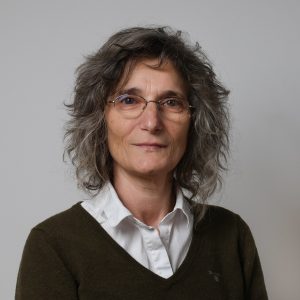}}]{Ib\'{e}ria Medeiros}
    is an Associate Professor at the Faculty of Sciences of the University of Lisbon (FCUL), and an integrated researcher of the LASIGE research unit. Her research interests are software security, vulnerability and attack detection, code privacy and correction, and artificial intelligence applied to cybersecurity. She is the author of tools for software security and cybersecurity, and has been involved in international and national projects. More information about her at \url{https://www.di.fc.ul.pt/~imedeiros/}.
\end{IEEEbiography}
%%
	
%\vfill
%\enlargethispage{-9.5cm}
\newpage

\end{document}